\documentclass{aa}
\usepackage{times}
\usepackage{graphics}
\begin{document}
\thesaurus{
06				
(03.20.1;
 08.03.1;
 08.03.4;
 08.09.2: IRC\,+10\,216;
 08.13.2;
 08.16.4)
}
\title{
The dynamical evolution of the fragmented, bipolar dust shell around the
carbon star IRC +10 216
\thanks{
based on observations performed with the 6~m telecope at
the Special Astrophysical Observatory, Russia
}
}
\subtitle{Rapid changes of a PPN-like structure?}
\author{
R.\ Osterbart\inst{1}\and
Y.\ Y.\ Balega\inst{2}\and
T.\ Bl\"ocker\inst{1}\and 
A.B.\ Men'shchikov\inst{1,3}\and
G.\ Weigelt\inst{1}
}
\institute{
Max--Planck--Institut f\"ur Radioastronomie, Auf dem H\"ugel 69,
53121 Bonn, Germany\\
(osterbart@mpifr-bonn.mpg.de,
bloecker@mpifr-bonn.mpg.de,
weigelt@mpifr-bonn.mpg.de)
\and
Special Astrophysical Observatory, Nizhnij Arkhyz,
Karachaevo--Cherkesia, 357147, Russia
(balega@sao.ru)
\and
Stockholm Observatory, 133 36 Saltsj\"obaden, Sweden
(sasha@astro.su.se)
}
\offprints{G.\ Weigelt}
%
\date{Received ~~~~ / Accepted ~~~~}
\titlerunning{The evolution of the dust shell of IRC +10\,216}
\authorrunning{Osterbart et al.}
\maketitle
\begin{abstract}
We present high--resolution $J$--, $H$--, and $K$--band observations and
the first $H-K$ color image of the carbon star \object{IRC +10 216}.
The images were reconstructed from 6 m telescope speckle interferograms
using the bispectrum speckle interferometry method.  The $H$ and $K$ images
with resolutions between 70~mas and 92~mas consist
of several compact components within a 0\farcs2 radius and a fainter
asymmetric nebula.  The brightest four components are denoted with A to
D in the order of decreasing brightness in the 1996 image.
A comparison of our images
from 1995, 1996, 1997, and 1998 gives --- almost like a movie of five frames
--- insight into the dynamical evolution of the inner nebula.
For instance, the separation of the two brightest components A and B
increased from 191 mas in 1995 to 265 mas in 1998.  At the same time,
component B is fading and the components C and D become brighter.
The {\sf X}--shaped bipolar structure of the nebula, most prominently
present in the $J$--band image, implies an asymmetric mass--loss.  Such
asymmetries are often present in protoplanetary nebulae but are
unexpected for AGB stars.  \object{IRC +10 216} is thus likely to be
very advanced in its AGB evolution, shortly before turning into a
protoplanetary nebula.
The cometary shapes of A in the $H$ and $J$ images and in the 0.79
$\mu$m and 1.06 $\mu$m HST images suggest that the core of A is not the
central star, but the southern lobe of a bipolar structure.  The
position of the central star is probably at or near the position of
component B, where the $H-K$ color has a value of 4.2.  If the star is
at or near B, then the components A, C, and D are likely to be located
at the inner boundary of the dust shell.

\end{abstract}
\section{Introduction}
\object{IRC +10 216} (\object{CW Leo}) is the nearest and best--studied
carbon star and one of the brightest infrared sources.  It experiences a
strong mass loss at a rate of $\dot{M} \approx
2-5\times10^{-5}$M$_{\odot}\,$yr$^{-1}$ (see e.g.\ Loup et~al.\
\cite{LoupForveilleEtAl93}).  The central star of \object{IRC +10 216}
is a long--period variable star (LPV) with a period of approximately
649\,days (Le~Bertre \cite{LeBertre92}).  Recent distance estimates of
110~pc to 135~pc (Groenewegen \cite{Groenewegen97}) and 150~pc (Crosas
\& Menten \cite{CrosasMenten97}) were reported.  \object{IRC +10 216}'s
initial mass can be expected to be close to 4\,M$_\odot$ (Guelin et~al.\
\cite{GuelinForestiniEtAl95}, Weigelt et~al.\
\cite{WeigeltBalegaEtAl98}).  The bipolar appearance of the nebula
around this object was already reported by Christou et~al.\
(\cite{ChristouRidgwayEtAl90}) and Kastner \& Weintraub
(\cite{KastnerWeintraub94}).  The non-spherical structure is consistent
with the conjecture that \object{IRC +10 216} is in a phase immediately
before entering the protoplanetary nebula stage.  High--resolution
observations of this object and its circumstellar dust shell were
reported by
McCarthy et~al.\ (\cite{McCarthyMcLeodEtAl90}),
Christou et~al.\ (\cite{ChristouRidgwayEtAl90}),
Dyck et~al.\ (\cite{DyckBensonEtAl91}),
Danchi et~al.\ (\cite{DanchiBesterEtAl94}),
Osterbart et~al.\ (\cite{OsterbartBalegaEtAl97}),
Weigelt et~al.\ (\cite{WeigeltBalegaEtAl97}, \cite{WeigeltBalegaEtAl98}
\cite{WeigeltBloeckerEtAl99}),
Skinner et~al.\ (\cite{SkinnerMeixnerEtAl98}),
and Haniff \& Buscher (\cite{HaniffBuscher98}).
The results of Dyck et~al.\ (\cite{DyckBensonEtAl91}) and Haniff \&
Buscher (\cite{HaniffBuscher98}) showed that the structure of the dust
shell of \object{IRC +10 216} has been changing for some years.
Detailed radiative transfer calculations for \object{IRC +10 216} were
recently performed by Ivezi\'{c} \& Elitzur (\cite{IvezicElitzur96}),
Crosas \& Menten (\cite{CrosasMenten97}), and Groenewegen
(\cite{Groenewegen97}) using a large amount of spectroscopic and
visibility data.
The aim of this paper is to discuss the properties of the inner dust shell of 
\object{IRC +10 216} on the basis of a series of high--resolution
observations. In a second paper (Men'shchikov et~al., in prep.)
we will present a detailed two--dimensional radiative transfer model for
this object.

\section{Observational results}
\begin{figure*}
  \setlength{\unitlength}{0.5mm}
  \begin{picture}(215,470)(0,0)
    \put(0,376){  \resizebox{46.5mm}{!}{\includegraphics{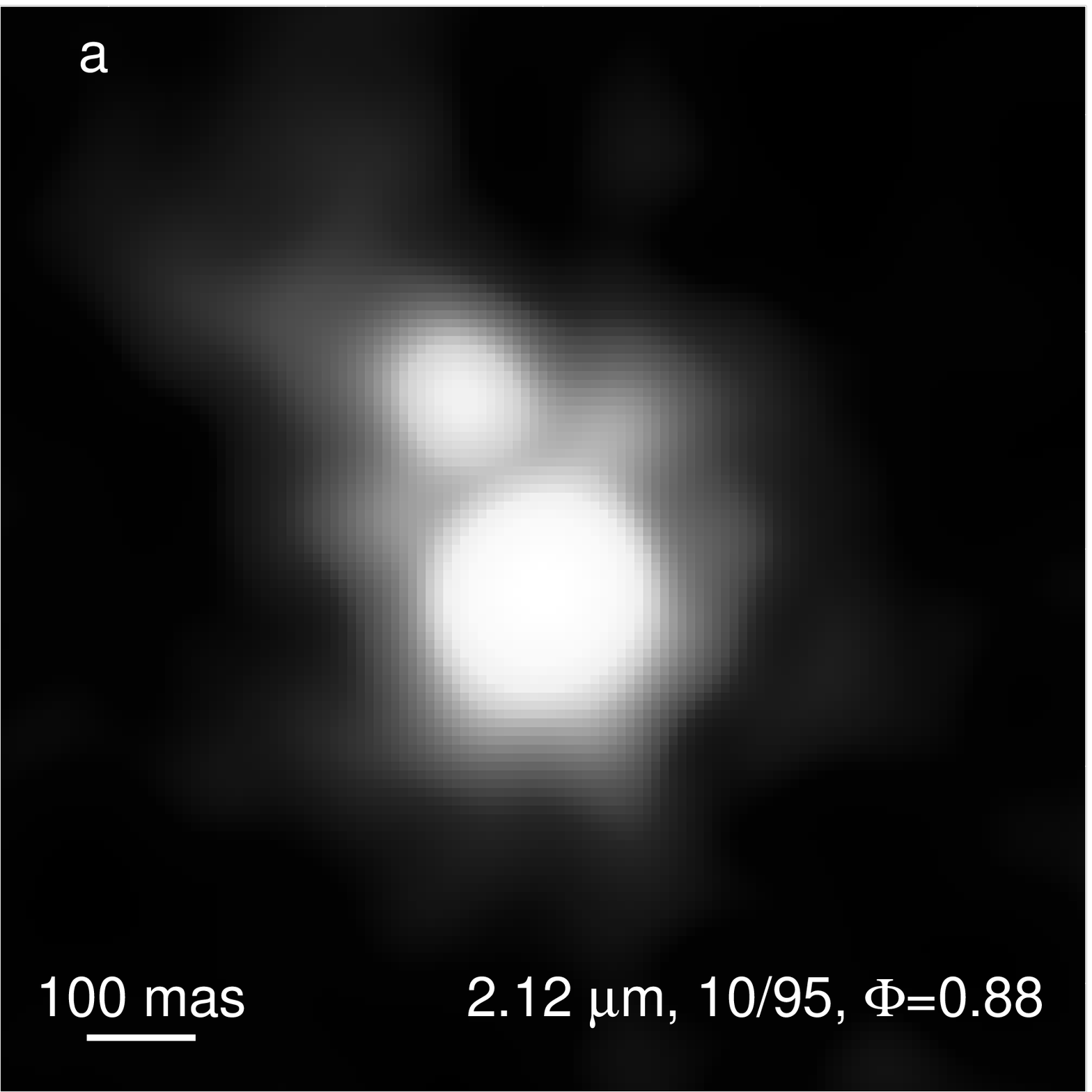}  } }
    \put(0,282){  \resizebox{46.5mm}{!}{\includegraphics{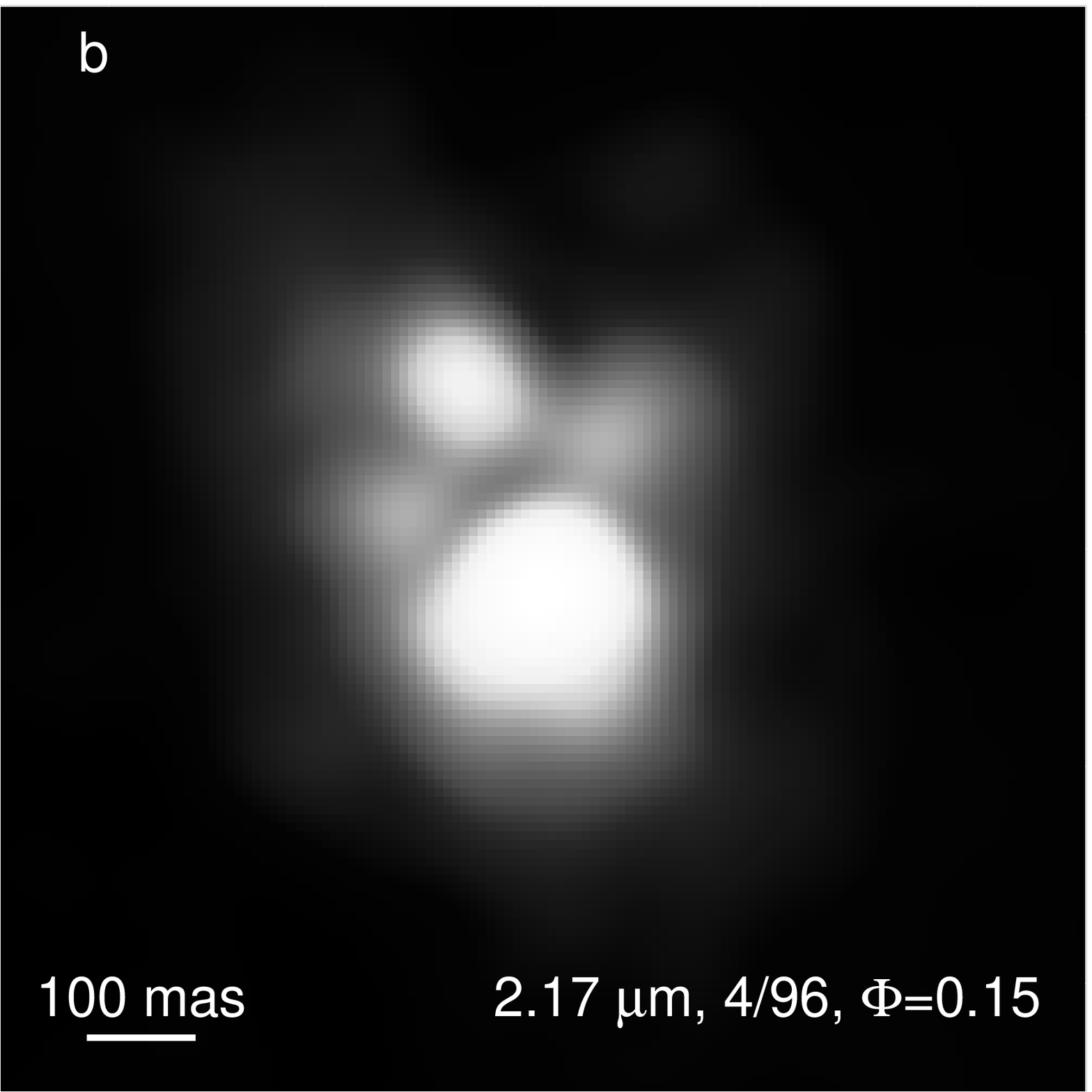}  } }
    \put(0,188){  \resizebox{46.5mm}{!}{\includegraphics{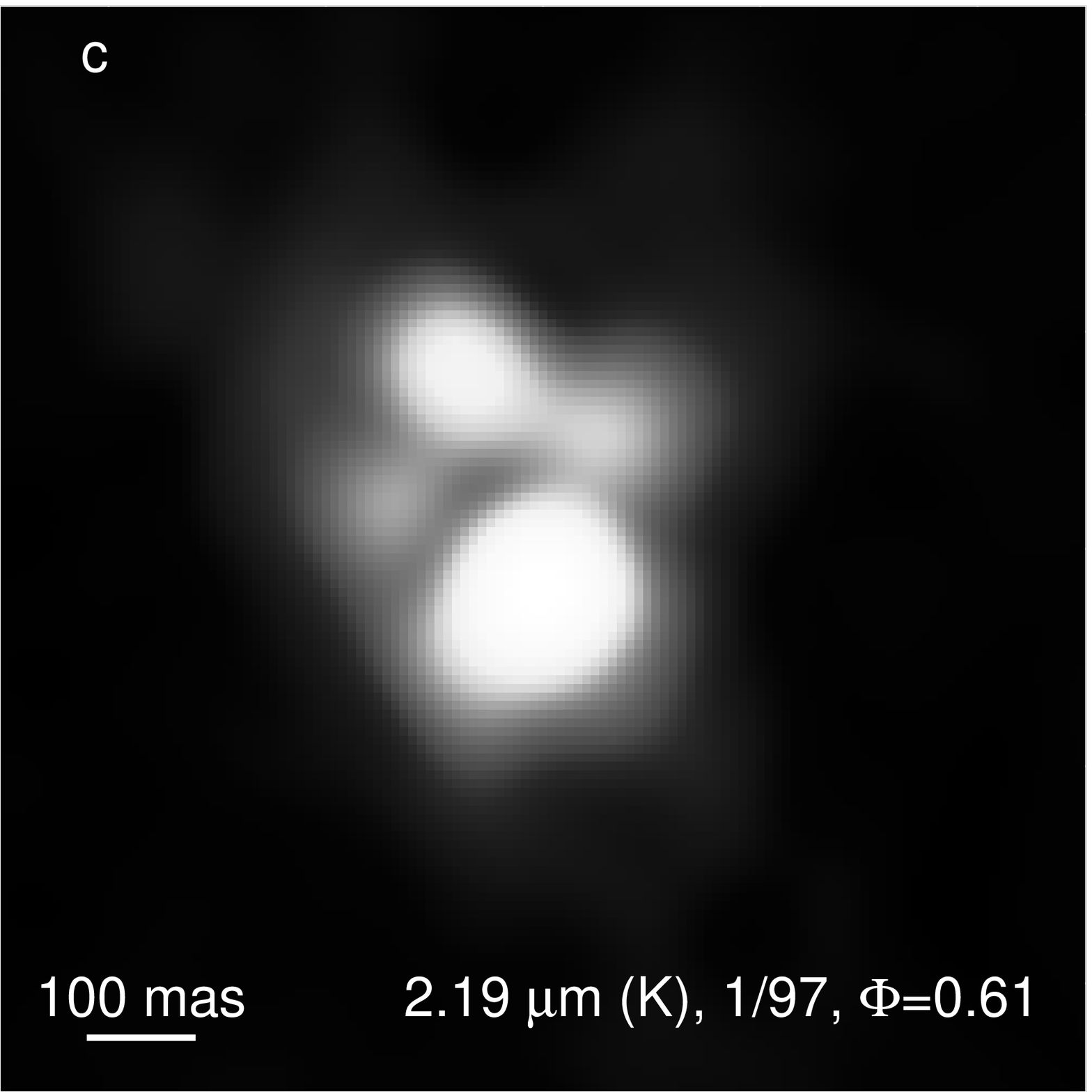}   } }
    \put(0,94){   \resizebox{46.5mm}{!}{\includegraphics{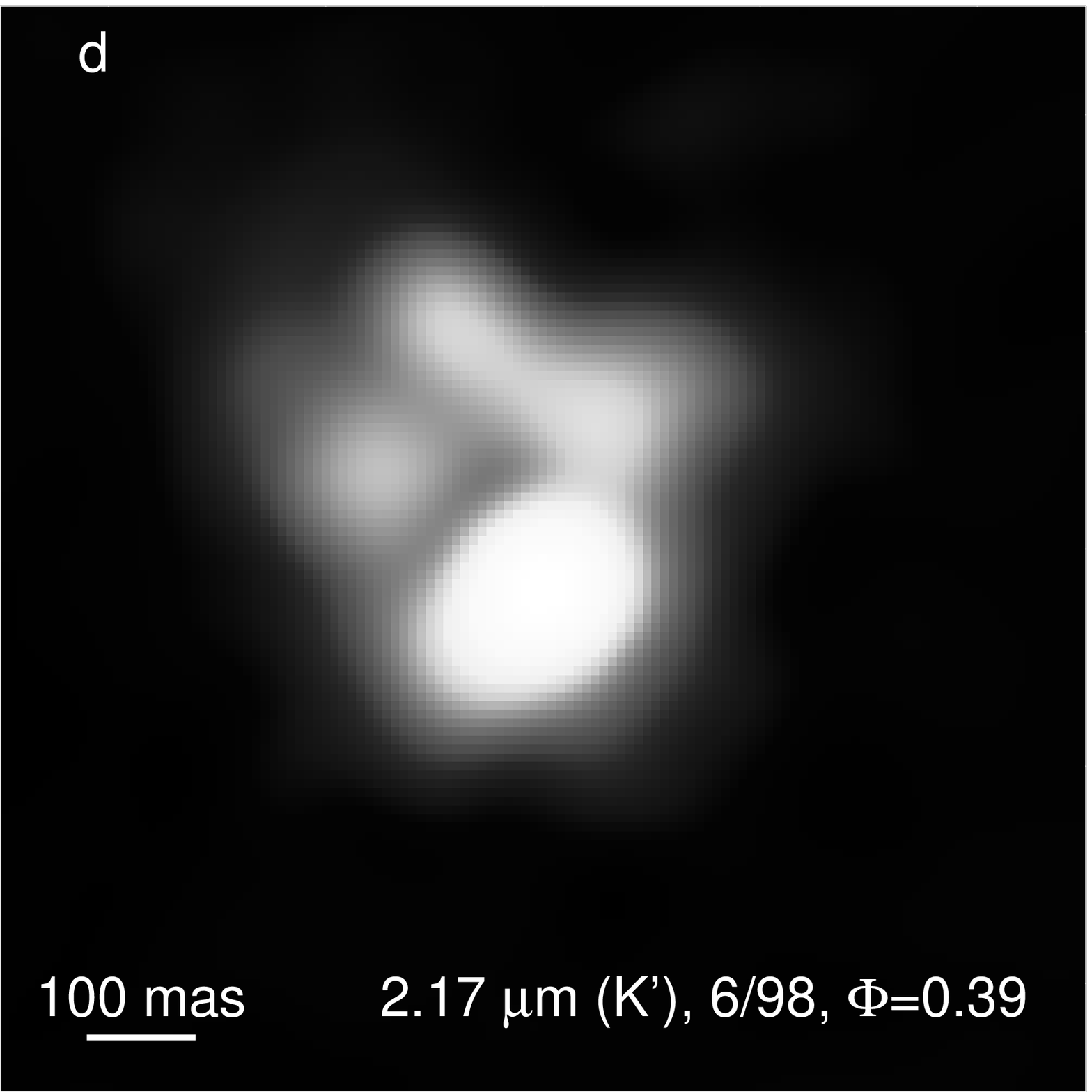}  } }
    \put(0,0){    \resizebox{46.5mm}{!}{\includegraphics{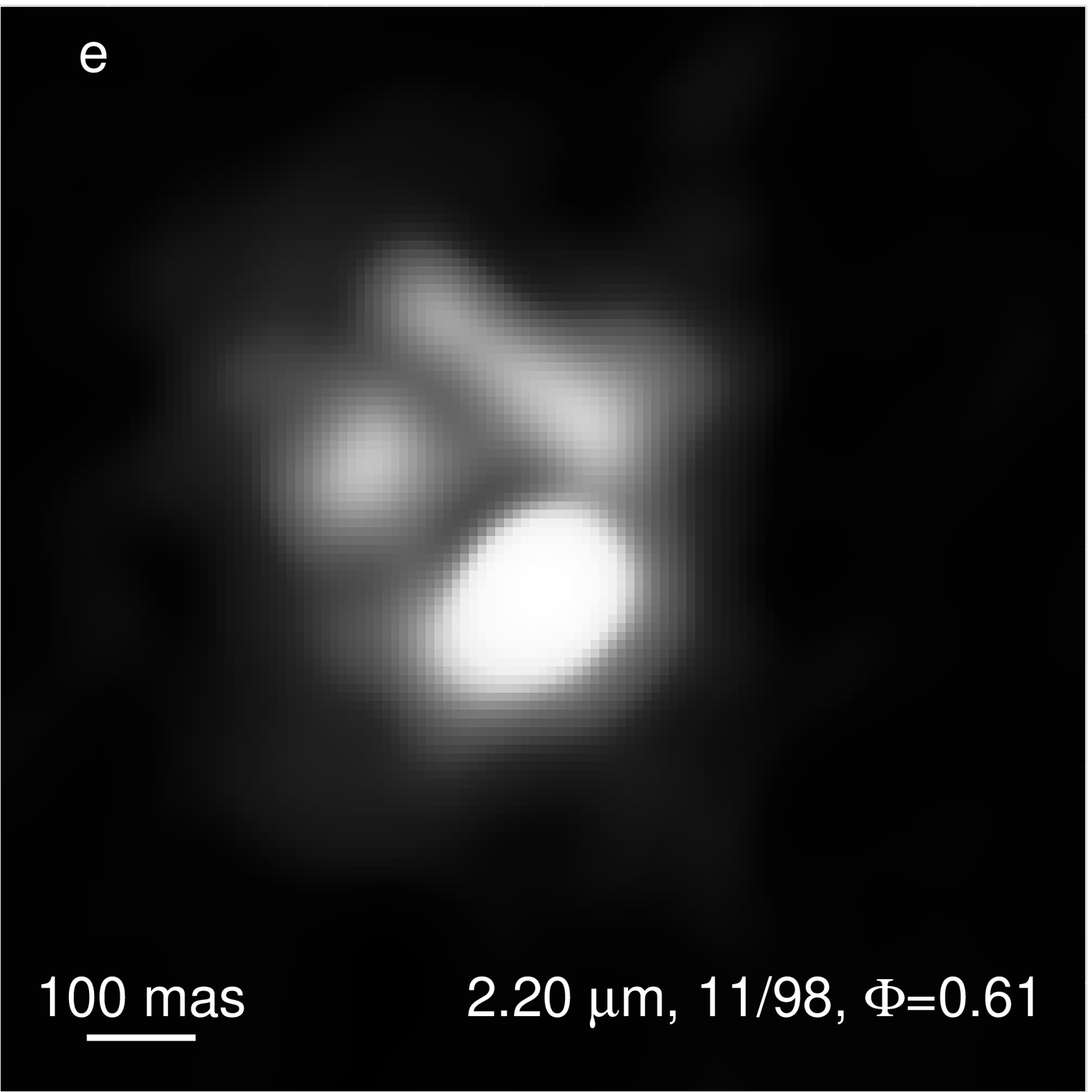} } }
    \put(94,376){ \resizebox{46.5mm}{!}{\includegraphics{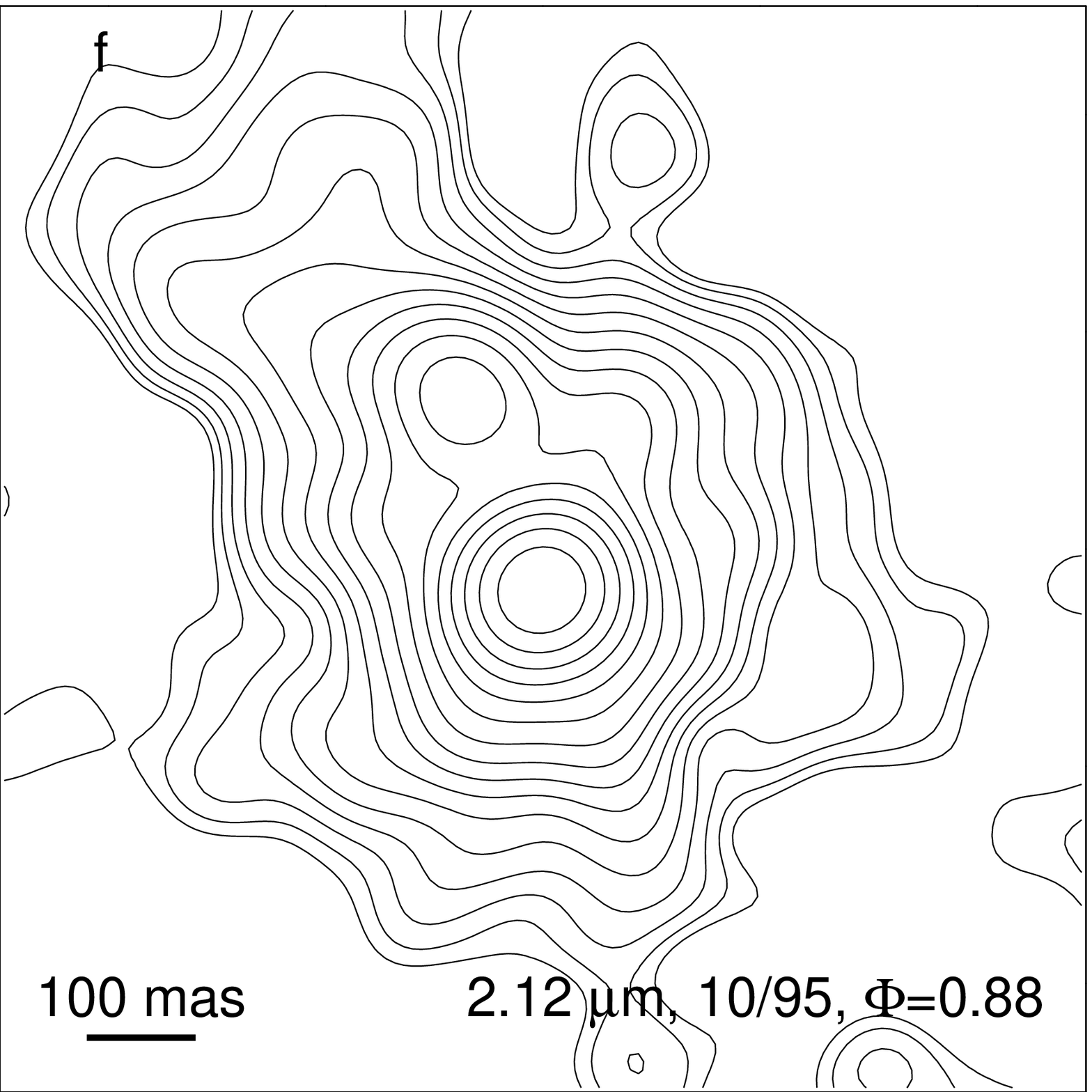}  } }
    \put(94,282){ \resizebox{46.5mm}{!}{\includegraphics{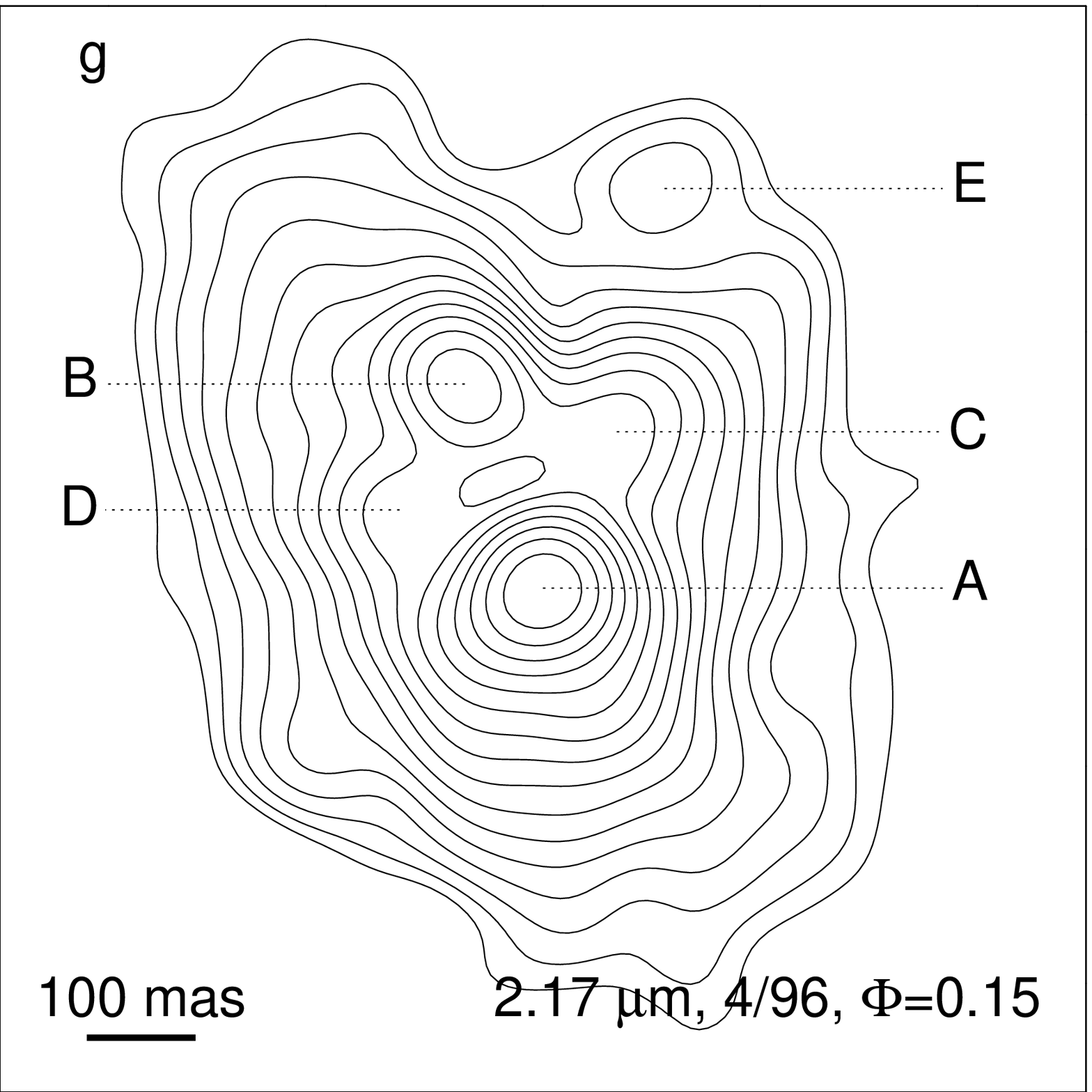}  } }
    \put(94,188){ \resizebox{46.5mm}{!}{\includegraphics{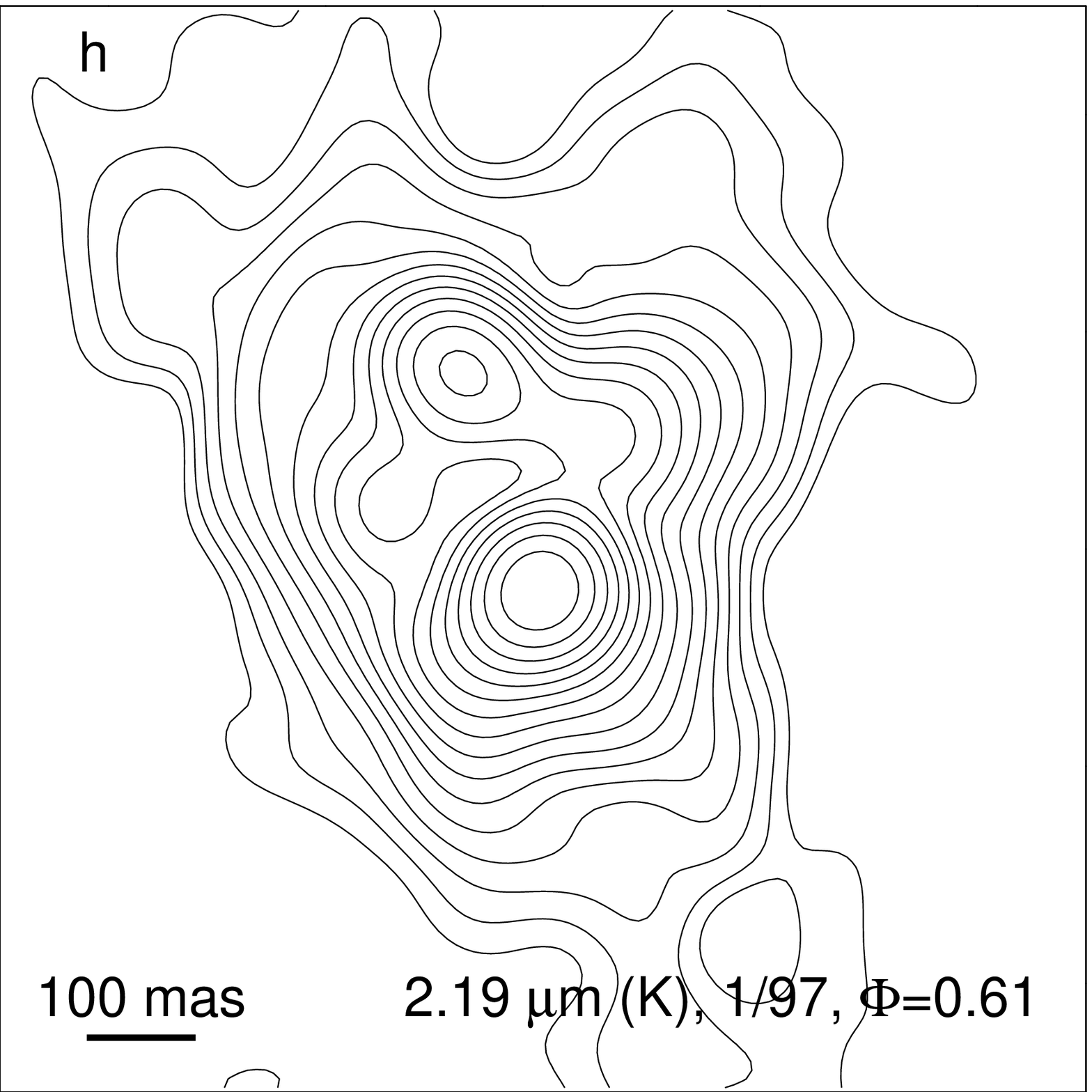}   } }
    \put(94,94){  \resizebox{46.5mm}{!}{\includegraphics{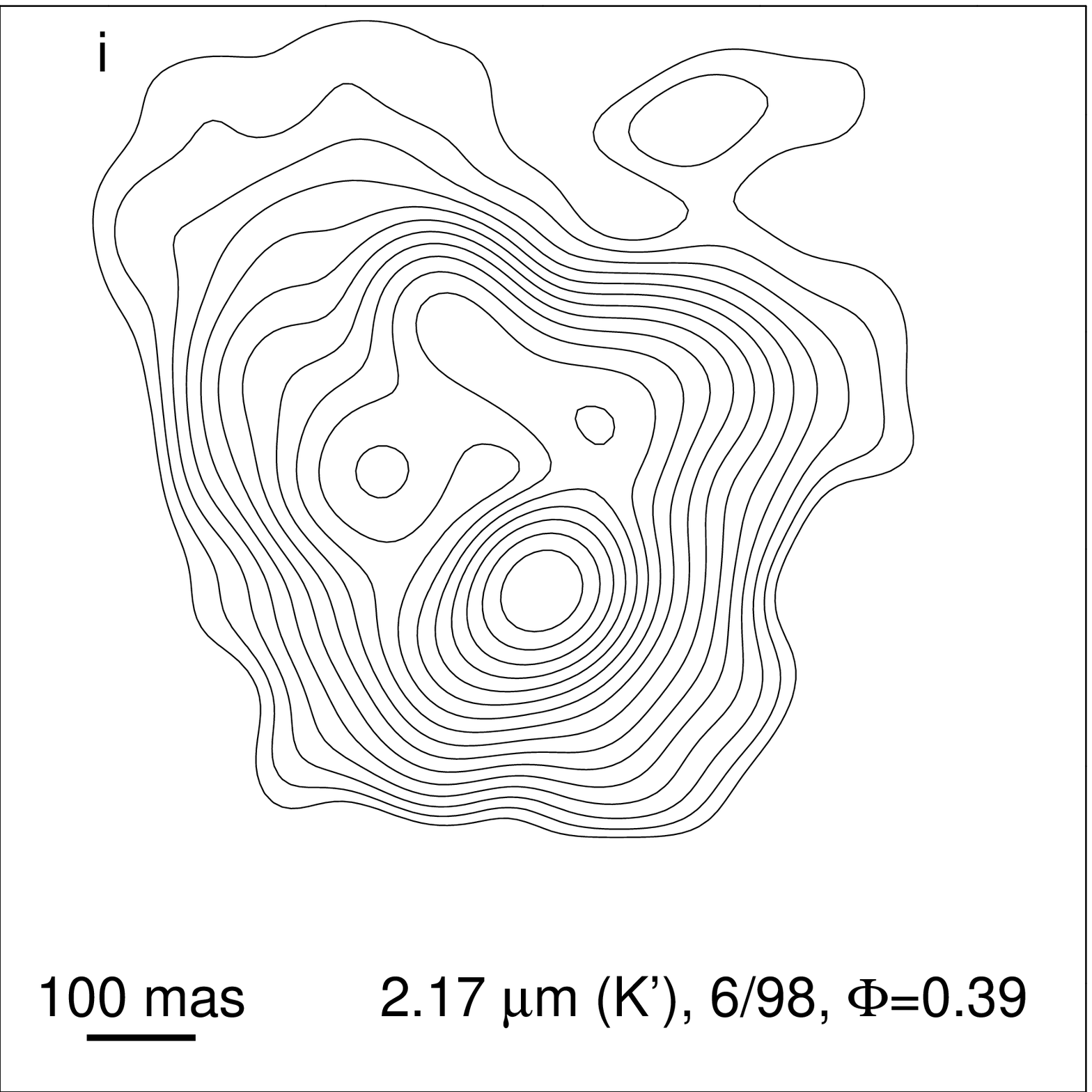}  } }
    \put(94,0){   \resizebox{46.5mm}{!}{\includegraphics{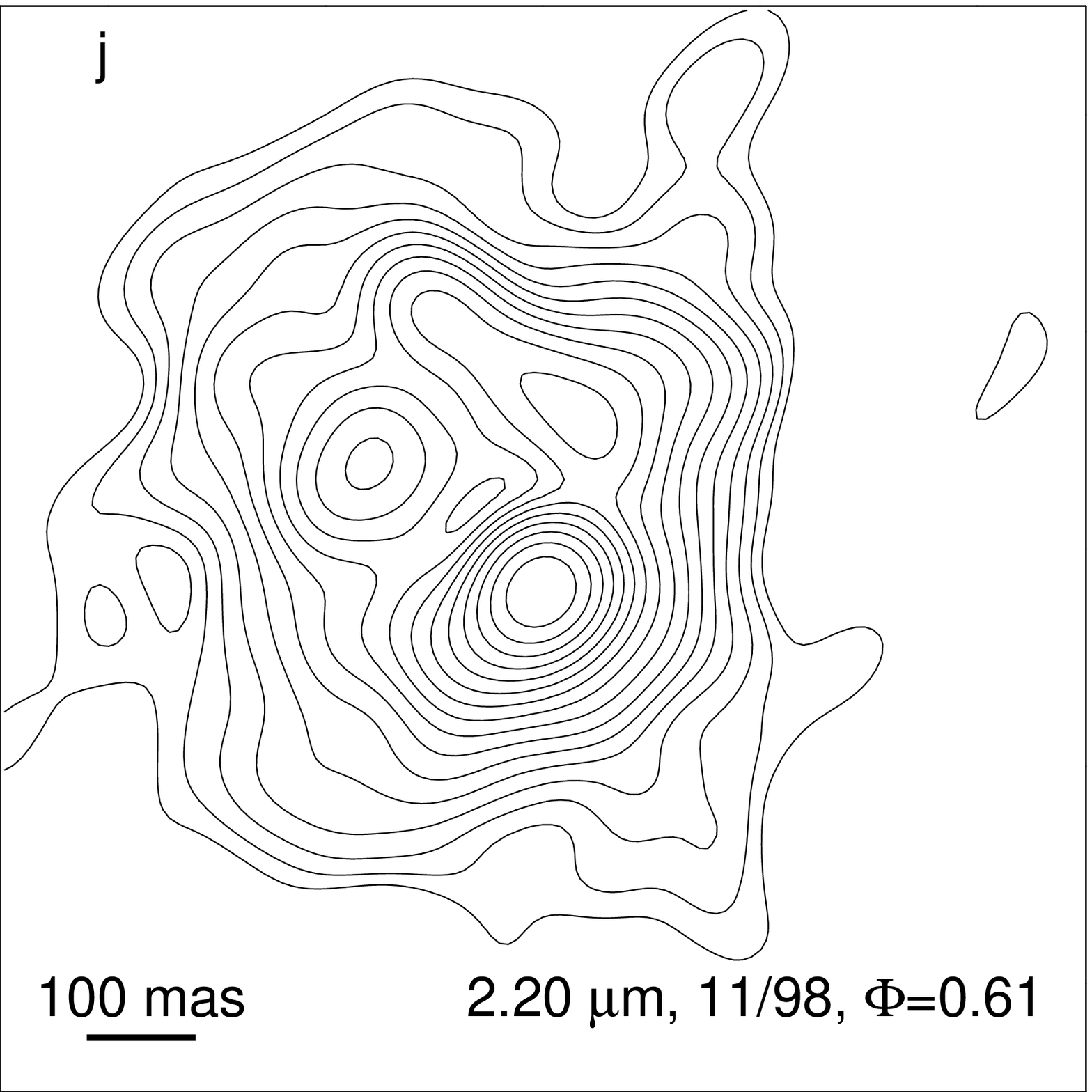} } }
    \put(213,151){  \resizebox{75mm}{!}{\includegraphics{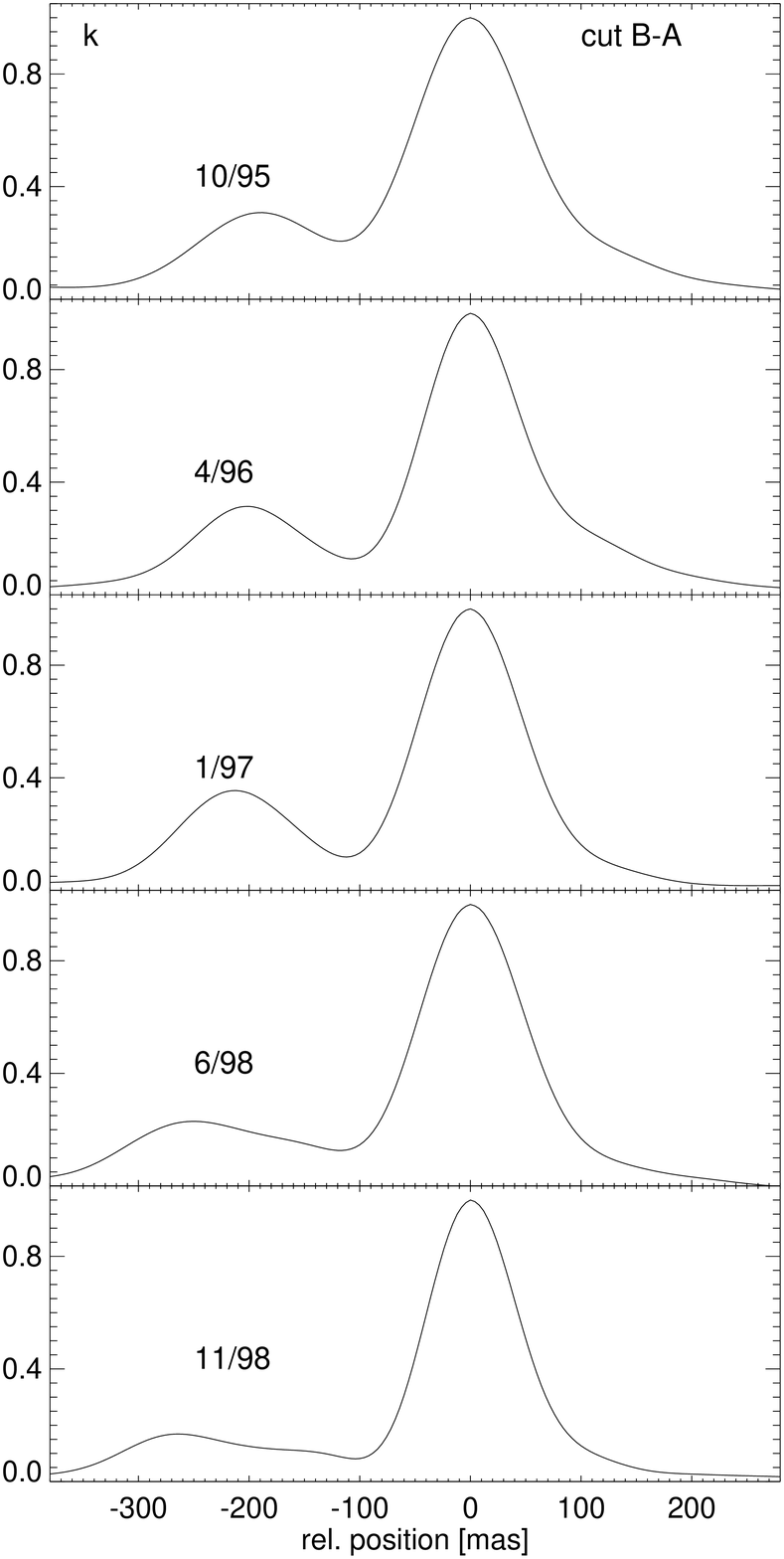} } }
  \end{picture}
  \hfill
  \parbox[b]{75mm}{
  \caption{
    {\bf a} to {\bf e}:
    High--resolution bispectrum speckle interferometry images
    of \object{IRC +10 216}.
    North is up and east to the left.  The figures represent a time
    series showing the evolution of the subarcsecond structure of
    \object{IRC +10 216} from 1995 (top) to 1998 (bottom).  In all
    figures the same gray level corresponds to the same relative
    intensity measured with respect to the peak.
    {\bf f} to {\bf j}:
    Same as a to e but as contour representation (contours at every 0\fm3
    down to 4\fm8 relative to the respective peak).  The resolutions of
    the images are 92~mas (a, f), 82~mas (b, g), 87~mas (c, h), 87~mas
    (d, i), and 75~mas (e, j).
    In panel g the denotation for the components A to E is shown.
    In all figures the epoch of the observation, the filter wavelength
    and the photometric phase $\Phi$ are indicated.
    {\bf k} Cuts through the images a to e along the axis from component A to
    B (position angle 20\degr)
    \label{recall}
  }
  }
\end{figure*}
\begin{table}
\caption{Observations.  The center wavelengths ($\lambda_\mathrm{c}$)
and FWHM bandwidths ($\Delta\lambda$) of the filters are given in
$\mu$m.  $N_\mathrm{T}$ and $N_\mathrm{R}$
are the numbers of target speckle interferograms for \object{IRC +10 216}
and reference star speckle interferograms
(\object{SAO 116569}, \object{SAO 116569},
\object{HIP 51133}, \object{HIP 47959},
\object{HIP 51133}, \object{HIP 52689},
\object{HIP 52689}, \object{HIP 50792}, \object{HIP 49583}
from top to bottom).  $T$ is the exposure time per frame in ms,
$S$ is the seeing, and $p$ is the pixel size in mas.
\label{obstab}}
\begin{tabular}{@{}ll@{\,}lrrrrr@{}}
\hline
Date & \multicolumn{2}{c}{Filter $\lambda_\mathrm{c}/\Delta\lambda$} &
$N_{\rm T}$ & $N_{\rm R}$ & $T$ & $S$ & $p$ \\
\hline
~8 Oct.\ 1995 &            & 2.17/0.02 &  216 &  253 & 100  & 1\farcs5 & 31.5 \\
~8 Oct.\ 1995 &            & 2.12/0.02 &  251 &  266 & 100  & 1\farcs5 & 31.5 \\
~2 Apr.\ 1996 & $J$        & 1.24/0.28 & 1196 &  981 & 200  & 1\farcs2 & 14.6 \\
~3 Apr.\ 1996 &            & 2.17/0.02 & 1112 & 1639 & 100  & 1\farcs6 & 30.6 \\
~3 Apr.\ 1996 & $K^\prime$ & 2.17/0.33 & 1403 & 1363 &  70  & 2\farcs5 & 14.6 \\
23 Jan.\ 1997 & $H$        & 1.64/0.31 & 1665 & 2110 & 100  & 1\farcs5 & 19.8 \\
23 Jan.\ 1997 & $K$        & 2.19/0.41 & 2165 & 1539 &  50  & 0\farcs9 & 30.6 \\
14 Jun.\ 1998 & $K^\prime$ & 2.17/0.33 &  800 &  571 &  50  & 1\farcs6 & 30.6 \\
~3 Nov.\ 1998 &            & 2.20/0.20 & 1087 &  842 &  40  & 1\farcs3 & 27.2 \\
\hline
\end{tabular}
\end{table}
The \object{IRC +10 216} speckle interferograms were obtained with the
6\,m telescope at the Special Astrophysical Observatory in Russia and
our {\sc Nicmos}\,3 camera at four epochs and with our {\sc
Hawaii}--array camera at one epoch (Nov.\ 1998).  Table~\ref{obstab}
lists the observational parameters.

\subsection{$J$-, $H$-, and $K$-band reconstructions}
\begin{figure*}
  \setlength{\unitlength}{0.5mm}
  \begin{picture}(360,140)(0,0)
    \put(0,0){    \resizebox{69.5mm}{!}{\includegraphics{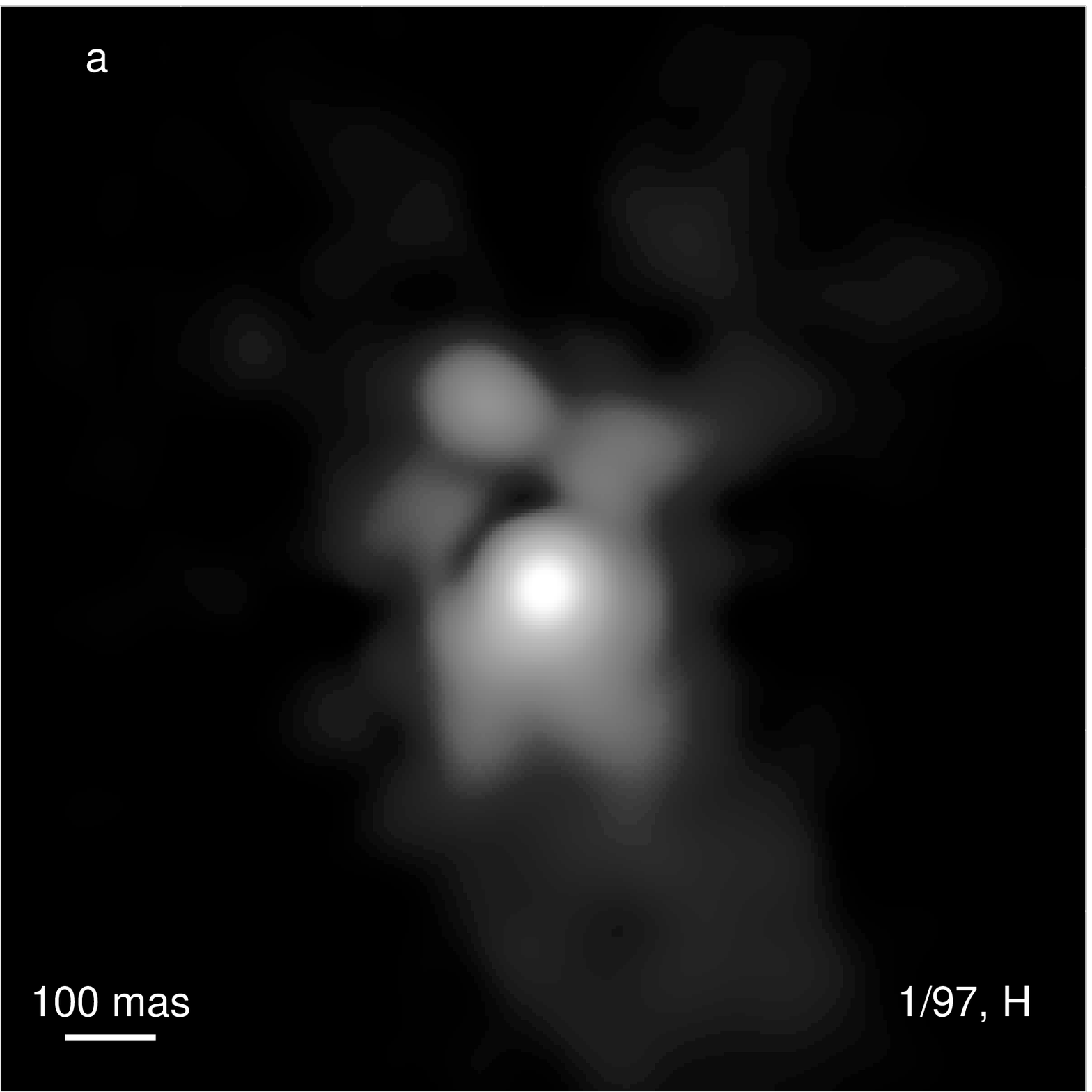}  } }
    \put(141,0){  \resizebox{69.5mm}{!}{\includegraphics{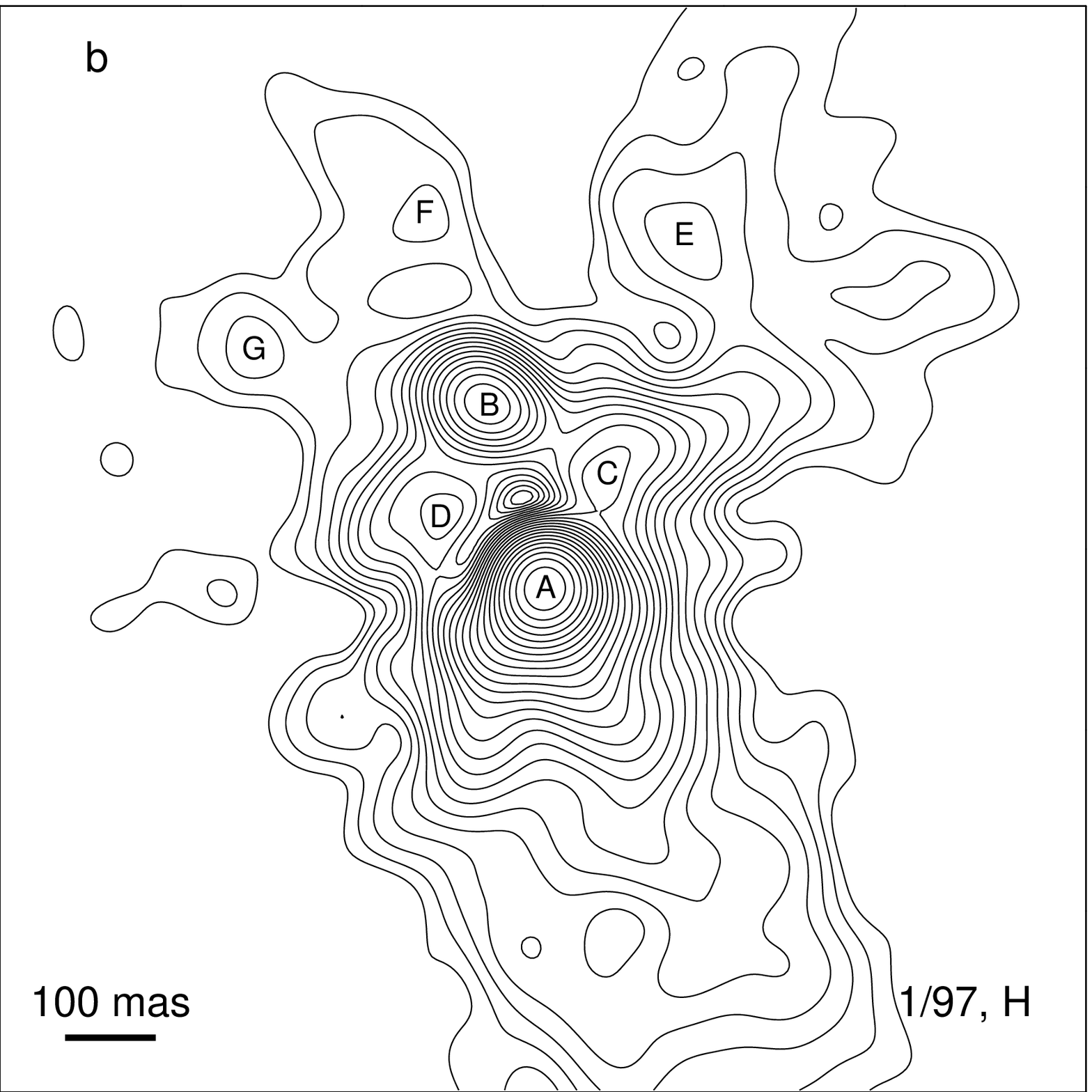}  } }
    \put(281,20){ \resizebox{39.5mm}{!}{\includegraphics{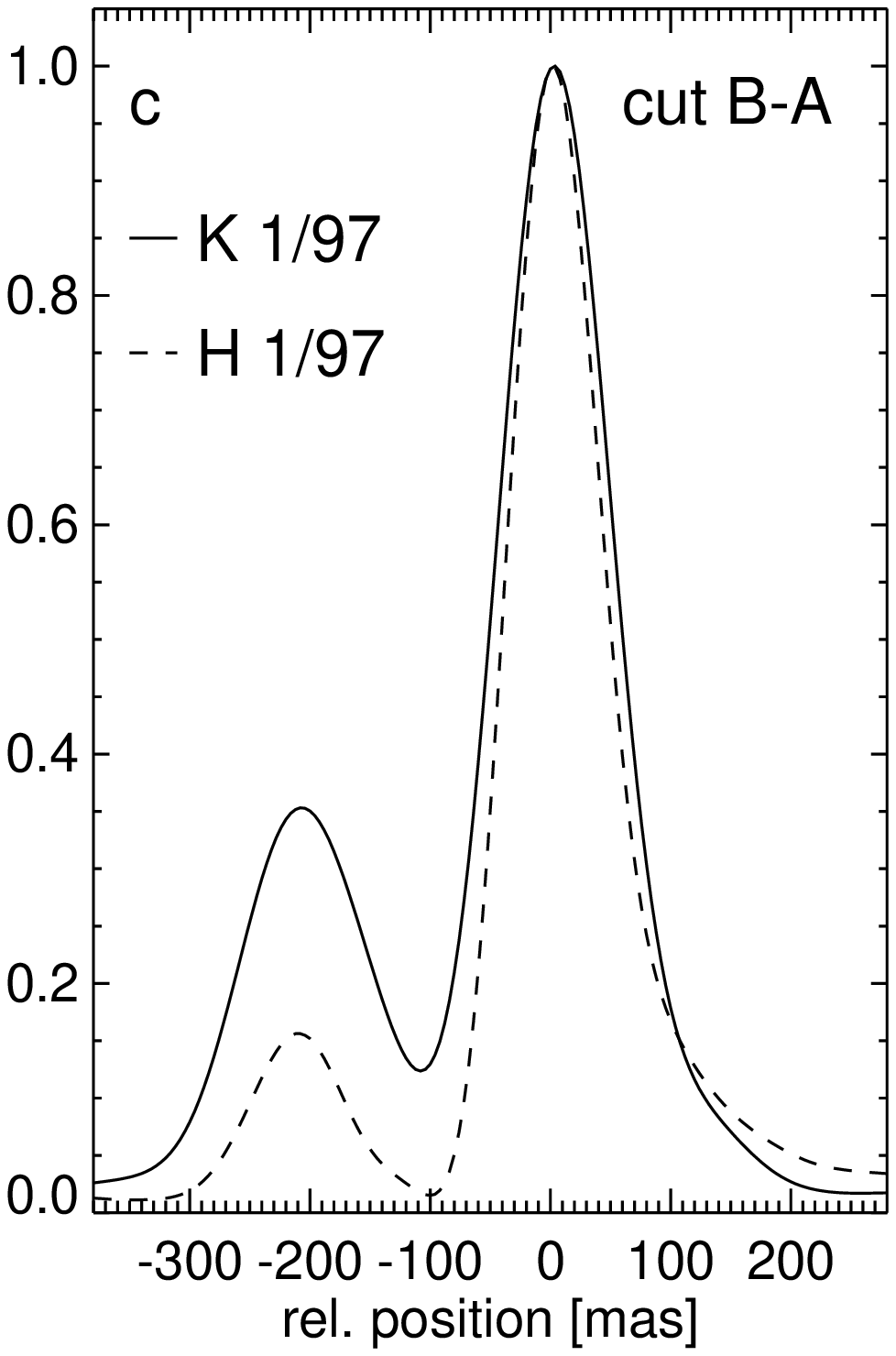}  } }
  \end{picture}
  \hfill
  \caption{
    {\bf a} 70~mas resolution bispectrum speckle interferometry image
    of \object{IRC +10 216} in the $H$--band.  North is up and east to
    the left.
    {\bf b} Same as {\bf a} as a contour image with denotations (A to G) of compact
    structures.  Contours are at 5\fm0 to 0\fm2 relative to the peak
    in steps of 0\fm2. 
    {\bf c} Normalized cuts through the $H$--band image {\bf a} and
    the $K$--band image of the same epoch (Fig.~\protect\ref{recall}c)
    in the direction of the components A and B (position angle
    $\sim$20\degr) illustrating the differences in the relative intensities
    of A and B for the different wavelengths
    \label{rech}
  }
\end{figure*}

Figs.~\ref{recall} and \ref{rech} show the $K$ and $H$ images,
respectively, of the central region of \object{IRC +10 216} for all
epochs.  The high--resolution images were reconstructed from the speckle
interferograms using the bispectrum speckle interferometry method
(Weigelt \cite{Weigelt77}, Lohmann et al.\ \cite{LohmannWeigeltEtAl83},
Weigelt \cite{Weigelt91}).  The achieved resolution of the images
depends slightly on the data quality (seeing and number of recorded
interferograms) and is indicated in the figure captions.  The object
power spectra were determined with the speckle interferometry method
(Labeyrie \cite{Labeyrie70}).  The speckle transfer functions were
derived from speckle interferograms of the unresolved stars mentioned in
Table~\ref{obstab}.

We denote the resolved components in the center of the nebula as A, B,
C, and D (see Figs.~\ref{recall}g and \ref{rech}b) in the order of decreasing peak intensity
(based on the $K$ band results from 1996).  In addition,
Fig.~\ref{rech}b shows three fainter components denoted with E, F, and G.  In
Fig.~\ref{rech}c, cuts through the centers of components A and B of the
$H$ and $K$ images from 1997 are shown to illustrate the differences in
the relative intensities of A and B for different wavelengths.

\begin{figure*}
  \setlength{\unitlength}{0.5mm}
  \begin{picture}(360,120)(0,0)
    \put(0,0){     \resizebox{59.5mm}{!}{\includegraphics{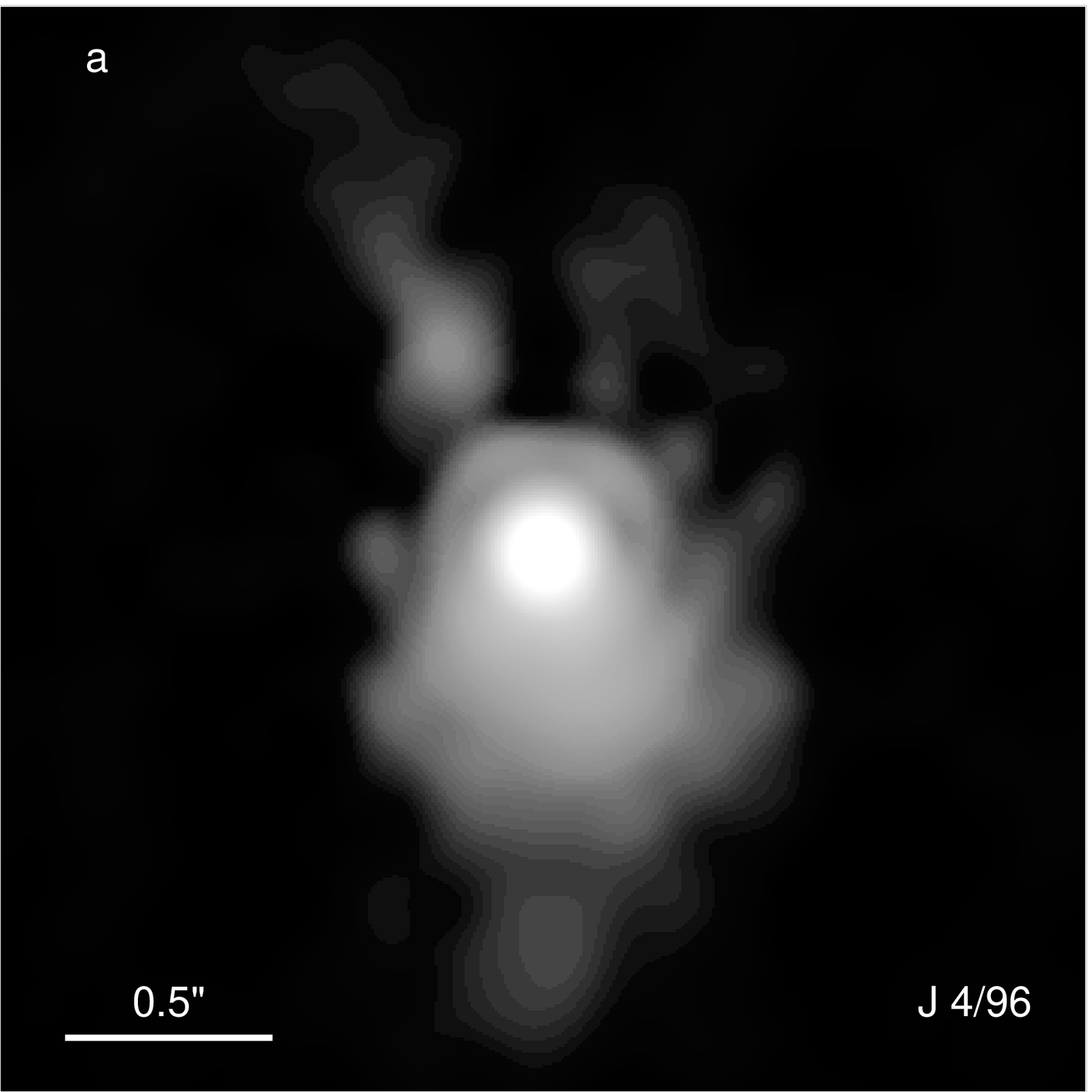}  } }
    \put(120,0){   \resizebox{59.5mm}{!}{\includegraphics{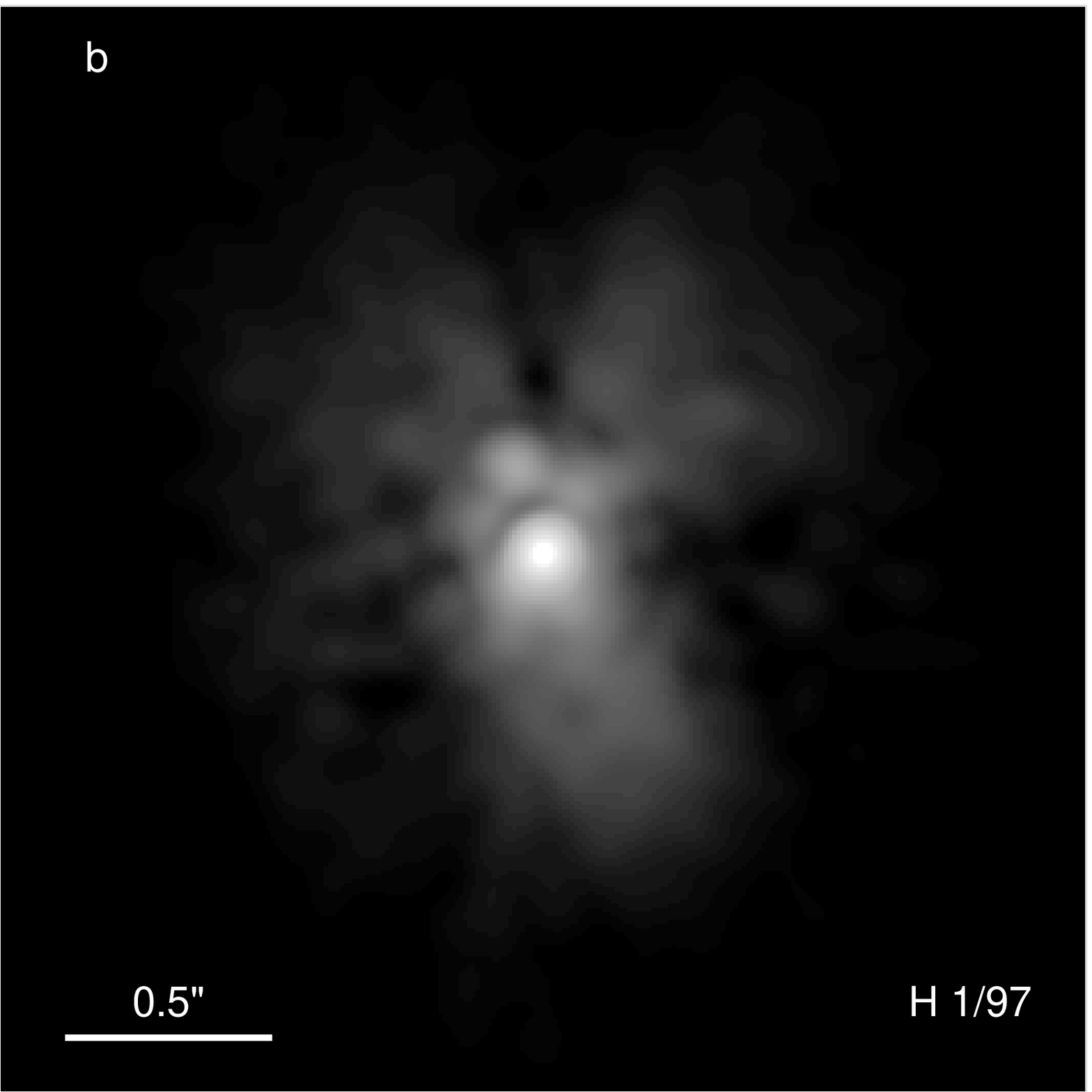}  } }
    \put(240,0){   \resizebox{59.5mm}{!}{\includegraphics{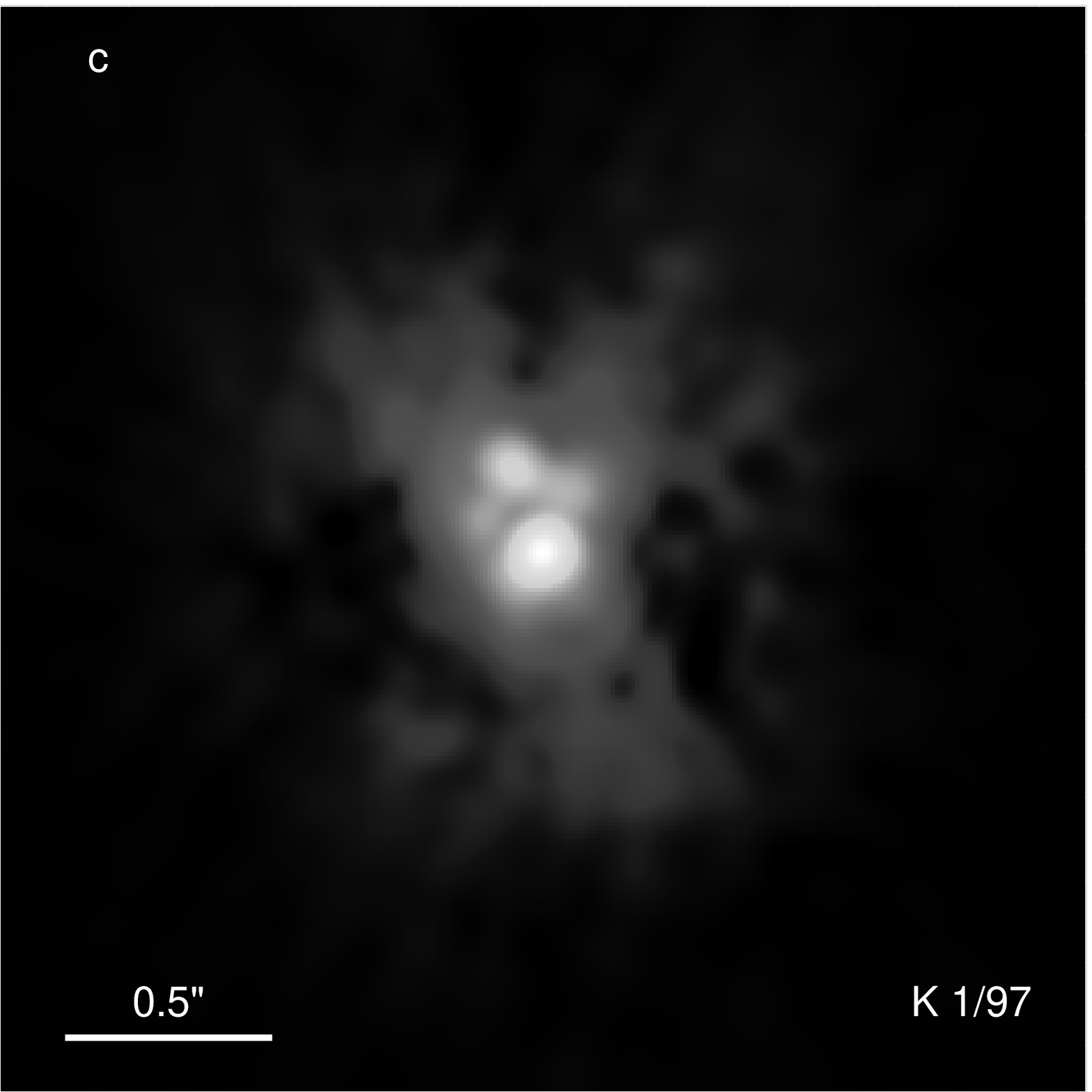}  } }
  \end{picture}
  \hfill
  \caption{
    {\bf a} $J$--band speckle reconstruction of \object{IRC +10 216} with
            149~mas resolution
	    showing the bipolar structure of the nebula.
    {\bf b} $H$--band speckle reconstruction of \object{IRC +10 216} with
            reduced resolution of 95~mas to increase the
            signal--to--noise ratio in the faint parts of the nebula.
    {\bf c} Same as {\bf b} for the $K$--band
    \label{recneb}
  }
\end{figure*}
Fig.~\ref{recneb} shows images of the faint nebula around
\object{IRC +10 216}, which seems to have a bipolar structure.  The
resolution of these images was reduced to 149~mas ($J$) and 95~mas
($H$ and $K$) to increase the signal--to--noise ratio in the
outer parts of the nebula.  The faint granular structure of the images
($\leq0.5\%$ of the peak intensity) is partly caused by speckle noise.

Note that the faint extended feature at position angle PA
$\sim$340\degr{} in the $J$ image corresponds quite well to a very faint
component (denoted with E in Figs.~\ref{recall}g and \ref{rech}b)
visible in all images in Figs.~\ref{recall} and \ref{rech} (assuming
that the brightest component in the $J$ image is roughly coinciding with
component A in the $H$ and $K$ images).  Two other faint features at PA
$\sim$20\degr{} and PA $\sim$50\degr{} (F and G in Fig.~\ref{rech}b) are
only visible in some of the images.

\subsection{Polarimetry}
\begin{figure*}
  \setlength{\unitlength}{0.5mm}
  \begin{picture}(360,120)(0,0)
    \put(0,0){     \resizebox{59.5mm}{!}{\includegraphics{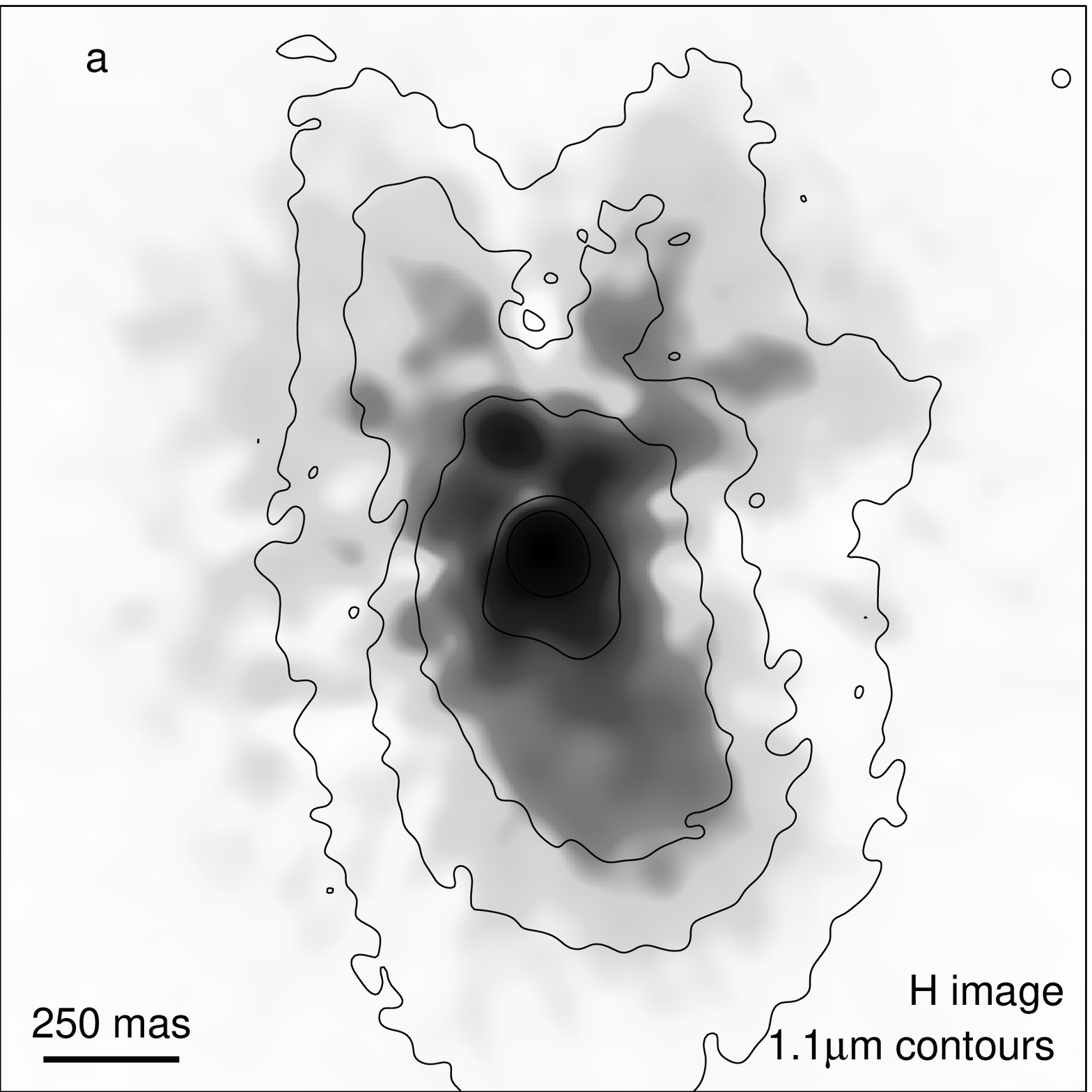}  } }
    \put(120,0){   \resizebox{59.5mm}{!}{\includegraphics{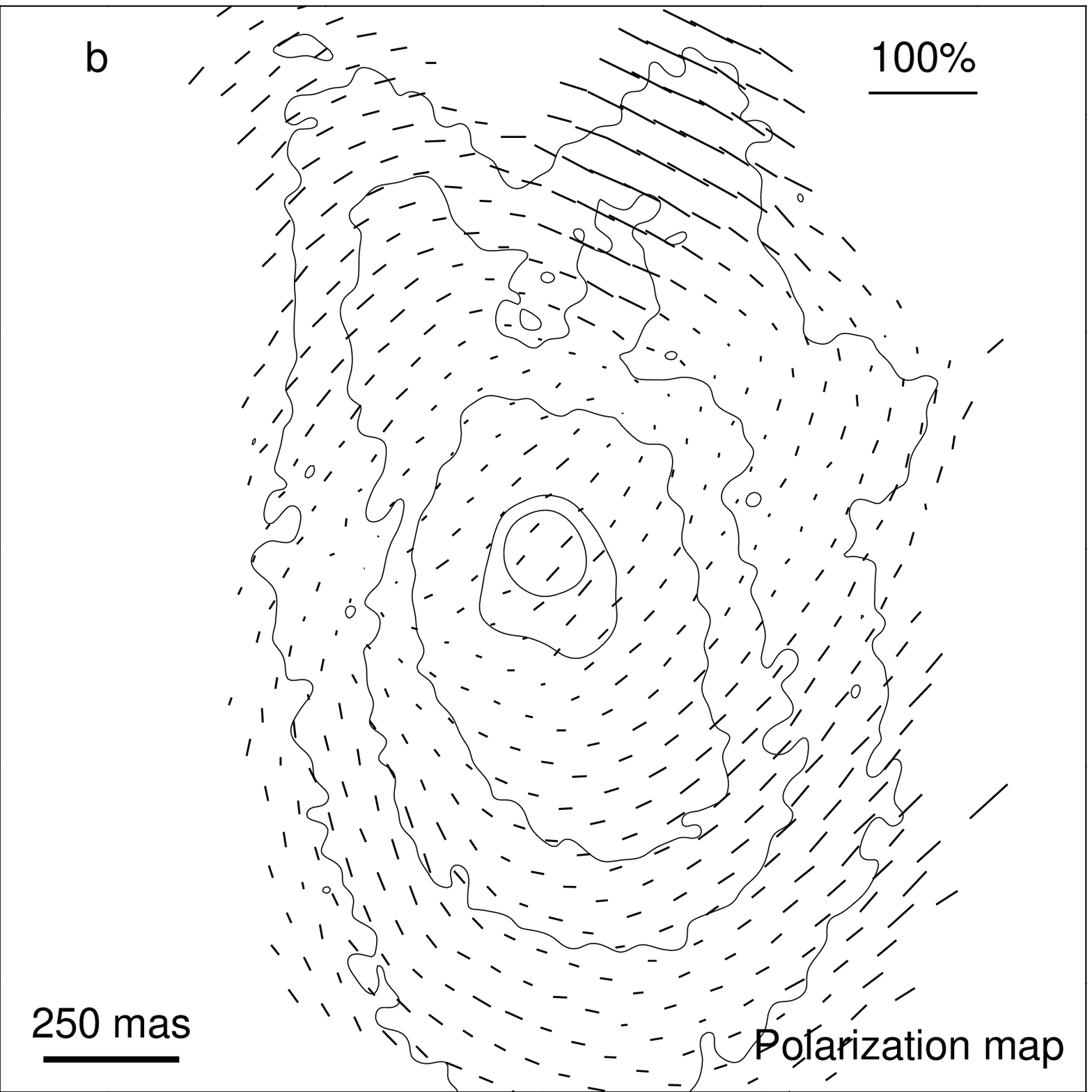}  } }
    \put(240,0){   \resizebox{59.5mm}{!}{\includegraphics{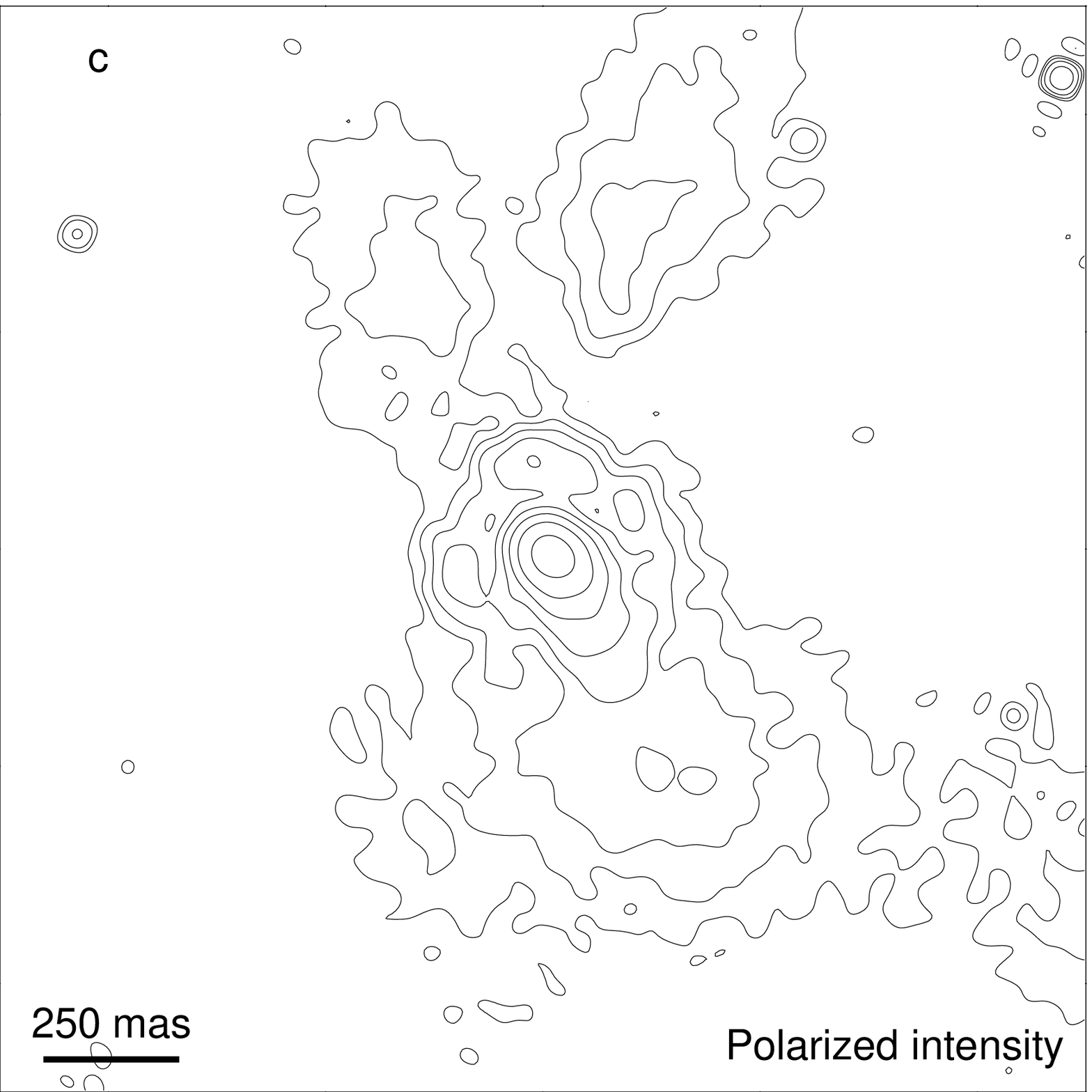}  } }
  \end{picture}
  \caption{
    {\bf a} Superposition of the total intensity at 1.1~$\mu$m
    (contours are at 5\fm0, 4\fm2, 3\fm4, 1\fm8, and 1\fm0 relative to
    the peak) derived from the archival HST polarization data and the
    negative $H$ speckle image.
    {\bf b} 1.1~$\mu$m polarization map.  The same contours as in
    frame {\bf a} are shown.
    {\bf c} Polarized intensity at 1.1~$\mu$m.  The contours are
    plotted from 2\fm4 to 6\fm0 difference relative to the peak of
    the total intensity in steps of 0\fm6
    \label{polmap}
  }
\end{figure*}
Fig.~\ref{polmap} shows the results of polarimetric observations with
the HST {\sc Nicmos} camera at a wavelength of 1.1~$\mu$m (raw data
retrieved from the Hubble Data Archive, STScI).  The data were obtained
on April 30, 1997 at a photometric phase of $\Phi=0.76$.  Data for three
polarization angles were available (0\degr, 120\degr, and 240\degr).  It
has been carefully checked that the three images were correctly
registered.  The images have been obtained without new pointing of the
telescope so that no correction was necessary (see Voit \cite{Voit97}).
From the data the total intensity (contours in Fig.~\ref{polmap}a and
b), the polarized intensity (Fig.~\ref{polmap}c), the degree and the
position angle of the polarization have been derived (see e.g.\ Voit
\cite{Voit97}, \hbox{Fischer} et~al.\ \cite{FischerHenningEtAl94}).
Fig.~\ref{polmap}b shows the derived polarization map superimposed on
the contours of the total intensity.  In Fig.~\ref{polmap}a the two
images were superimposed by centering the brightest peaks onto each
other.  Looking at the southern tails of A and at the northern arms this
centering seems to be appropriate.  The structure of the 1.1~$\mu$m HST
image at the position of B is affected by the diffraction pattern of A.
Nevertheless, the contours indicate that a structure corresponding to B
might be present in the 1.1~$\mu$m image.
\subsection{The core structure and the faint nebula}
\label{obsneb}
The multi--component structure of the \object{IRC +10 216} $K$--band
image (of the innermost 300~mas~$\times$~300~mas) has already been
reported by Weigelt et~al.\ (\cite{WeigeltBalegaEtAl98}) and Haniff \&
Buscher (\cite{HaniffBuscher98}).  This bright inner region is
surrounded by a larger faint nebula (with $\sim1\%$ of the peak
brightness of A; see Fig.~\ref{recneb}b and c).
Three arms of the nebula can be seen at position angles of
$\sim$30\degr{} (NE), 340\degr{} (NW), and 210\degr{} (SW) with respect
to component A.  At 160\degr{} (counterside of the NW-arm) the nebula is
much fainter.  The direction from component A to component B
(PA$\sim$20\degr{}) can be taken as the direction of the main
axis (see also Kastner \& Weintraub \cite{KastnerWeintraub94}).
This direction fits well to the main axis of the H$^{13}$CN($J=1-0$)
emission (Dayal \& Bieging \cite{DayalBieging95}) which is weakly
elongated on a scale of about 10$^{\prime\prime}$. On larger scales
asymmetries are not observed (Dayal \& Bieging \cite{DayalBieging95},
Groenewegen et~al.\ \cite{GroenewegenVanDerVeenEtAl97}, Mauron \& Huggins
\cite{MauronHuggins99}).

\subsection{The bipolar structure in $J$ and at shorter wavelengths}
\label{jbipolar}
The $J$--band image (Fig.~\ref{recneb}a) and the 0.79 $\mu$m and 1.06
$\mu$m HST images (Haniff \& Buscher \cite{HaniffBuscher98}, Skinner
et~al.\ \cite{SkinnerMeixnerEtAl98}, see also Fig.~\ref{polmap}a for the
total intensity of the HST 1.1~$\mu$m polarization data) show a bipolar
shape of the nebula.  The southern lobe has a cometary or fan--shaped
structure whereas the northern area of the images shows two arms
reminiscent to (but weaker than) the northern {\sf X}--shaped structure
of the Red Rectangle (see, e.g., Men'shchikov et~al.\
\cite{MenshchikovBalegaEtAl98}).  However, the fact that even in the
polarized intensity the nebula is very faint on the southeastern side
suggests the main axis to be at PA$\sim$20\degr{} to 30\degr.

\subsection{$H-K$ color image}
\label{obscolor}
For the January 1997 data, the integral intensities in the full fields of view
of our camera (5\farcs1 in $H$ and 7\farcs8 in $K$) were compared to those of
the photometric standard stars \object{HIP 71284} (\object{BS 5447}) and
\object{HD 106965} (cf.\ Elias et~al.\ \cite{EliasFrogelEtAl82}).  The
resulting \object{IRC +10 216} magnitudes are $K=2.3$ and $H=5.4$.
According to an extrapolation of the fitted light curve of Le~Bertre
(\cite{LeBertre92}), \object{IRC +10 216} was very close to its light
minimum in January 1997.  Our magnitudes are in good agreement with this
prediction.  In a square aperture of 1\farcs6 the magnitudes can be
determined to be $K=2.5\pm0.1$ and $H=5.7\pm0.1$.
The integral color in this field is
thus $H-K\approx3.2$.  The resolved two--dimensional $H-K$ color image
is shown in Fig.~\ref{hkcolor} together with the contours of the $H$--band
image.
The $H$ and $K$ images used for the determination of the $H-K$ color
were reconstructed with a common resolution of 95~mas which is also the
resolution of the color image.
The dependence of the $H-K$ color image on relative shifts
between the two images was investigated because it is not a priori known
whether
the positions of the intensity maxima in $H$ and $K$ coincide.  The relative
position of the components A and B is very similar in the $H$ and $K$ images
from 1997 so that a solution could be found where both
components A and B are almost coinciding (within a few mas)
for the two images.  We found that the color image is not very
sensitive to relative shifts within the realistic uncertainties of some
milli--arcseconds.

\subsection{Separation of components A and B}
\label{obsmotion}
\begin{figure}
  \resizebox{88mm}{!}{\includegraphics{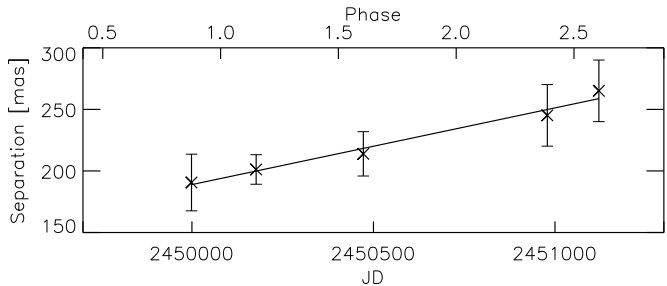} }
  \caption{
    Separation of the components A and B (see Sect.~\ref{obsmotion}) as a function of
    time and photometric phase
    \label{moves}
  }
\end{figure}

Fig.~\ref{moves} shows the separation of the components A and B in the
$K$--band images for different epochs.  Phase 0 in Fig.~\ref{moves}
corresponds to JD=2449430.  This date of the photometric maximum was
derived from the results of Le~Bertre (\cite{LeBertre92}).  The
separations
are:
191~mas,
201~mas,
214~mas,
245~mas, and
265~mas,
for the 5 epochs from 1995 to 1998 shown in Fig.~\ref{recall}.
The linear regression fit gives a value of
23~mas/yr for the average increase in the apparent separation of the
components.  Interpreting this increase as a real motion would lead to
14~km/s within the plane of the sky (for a distance of $D=130$~pc).

The apparent relative motion of the nebula components is obviously not
simply related to the stellar variability which has a period of approximately
649~days (Le~Bertre \cite{LeBertre92}).  It may thus be related to
either an overall expansion of an inhomogeneous circumstellar dust
medium or a variability of the dust shell with a period significantly
larger than the stellar pulsation period (cf.\ Fleischer et~al.\
\cite{FleischerGaugerEtAl92}, Winters et~al.\
\cite{WintersFleischerEtAl94}, \cite{WintersFleischerEtAl95}).

Table \ref{allmotions} lists the apparent relative velocities within the
plane of the sky of the components A to E with respect to either
A or B or the center positions defined by A and B or by A, B, C, and D.
The values were determined from linear regression fits.

\begin{table}
\caption{
  Apparent velocities of components A, B, C, D, and E with respect to
  A, B, the center between A and B, and the center between A, B, C, and D,
  respectively.
  \label{allmotions}
  }
\begin{tabular}{lr@{~}rr@{~}rr@{~}rr@{~}r}
\hline
 & \multicolumn{2}{c}{~~A~~} & \multicolumn{2}{c}{~~B~~} &
   \multicolumn{2}{c}{center} & \multicolumn{2}{c}{center} \\
 & \multicolumn{2}{c}{~} & \multicolumn{2}{c}{~} &
   \multicolumn{2}{c}{AB} & \multicolumn{2}{c}{ABCD} \\
\hline
\multicolumn{9}{c}{mas/yr (km/s at $D=130$~pc)} \\
\hline
A &  -&     & 23&(14) & 11&(7)  & 14&(9)  \\
B & 23&(14) &  -&     & 11&(7)  & 10&(6)  \\
C &  8&(5)  &  7&(5)  & -2&(-1) &  0&(0)  \\
D & 21&(13) &  7&(5)  &  9&~(5) &  6&(4)  \\
E & 39&(24) & 25&(15) & 31&(19) & 31&(19) \\
\hline
\end{tabular}
\end{table}
\subsection{Structural changes within the nebula.}
The images by Haniff \& Buscher (\cite{HaniffBuscher98}) taken in 1989
and 1997, as well as the one--dimensional data collected by Dyck et~al.\
(\cite{DyckBensonEtAl91}) covering a larger range of epochs, already
show that
the structure of the envelope changes on time scales of some
years.  Our $K$--band observations now allow us to
study the changes within approximately 2 stellar pulsation cycles in more
detail.  Besides the apparent motion of the components, Fig.~\ref{recall} shows
that these components change their shapes and relative fluxes.  The
brightest component A appears to be rather symmetric in 1995.  At the
later epochs it approximately keeps its size perpendicular to the axis
A--B ($\sim$20\degr) but becomes narrower along this axis.  The
peak--to--peak intensity ratio of B and A is approximately constant from 1995
to 1997. Thereafter the component B started fading. At the same time the other
components become brighter and detached from A.  Note that the
photometric phases of \object{IRC +10 216} in January 1997 and November
1998 are almost identical. In fact, the integral $K$ magnitudes were
the same ($K=2.3$).  Again we find that the time scale
for the changes seen in our images is significantly different from the
period of the stellar pulsation.

\begin{figure}
  \centering
  \resizebox{50mm}{!}{\includegraphics{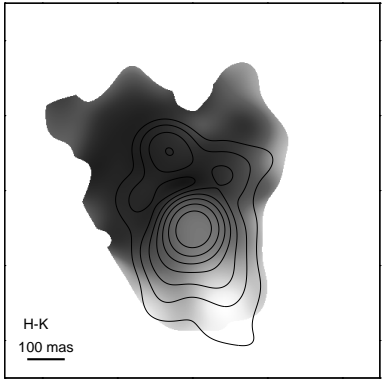} }
  \caption{
    $H-K$ color image of the central region of \object{IRC +10 216} in
    January 1997.  Darkest gray corresponds to $H-K=4.6$ and white
    corresponds to $H-K=2.1$. The color at the position of the brightest
    component (A) is $H-K=3.2$. The contours are those of the $H$--band
    image with a spacing of 0\fm5
    \label{hkcolor}
  }
\end{figure}
\section{Discussion}
\label{discussion}
The observations presented in the previous section as well as other
high--resolution observations (e.g.\ Haniff \& Buscher
\cite{HaniffBuscher98}) provide detailed information on the variable
structure in \object{IRC +10 216}.  The question of where, behind all the
dust, the central star is located, is of specific interest to
understand the physical properties of the nebula.

At short wavelengths the nebula shows a bipolar structure.  A comparison
of the observed structure with other bipolar nebulae like the Red
Rectangle (Men'shchikov et~al.\ \cite{MenshchikovBalegaEtAl98}) suggests
that the {\sf X}--like arms originate mainly from scattering of stellar
light on the surfaces of cavities.  The star is then at least partially
obscured by an optically thick dust shell or torus.  Haniff
\& Buscher (\cite{HaniffBuscher98}) argued convincingly
that the main axis of the
object is tilted with its southern side towards the observer.

\subsection{Is component A the star?}
\label{disa}
{\em Motion of the components.}
From molecular line observations the terminal velocity of the expansion
of the circumstellar envelope was found to be $\sim14$ to 15~km/s
(Dayal \& Bieging \cite{DayalBieging95}, Gensheimer \& Snyder
\cite{GensheimerSnyder97}).  The velocities of B and D (14~km/s and
13~km/s, resp.)  with respect to A can, therefore, only be understood if
the direction of motion is approximately within the plane of the sky.
However, the structure of the bipolar nebula at short wavelengths shows
that we are looking neither pole--on nor edge--on at the nebula but at
an intermediate viewing angle (e.g.\ 50\degr{} to 60\degr{}).  The
assumption that A is the star is thus not satisfactorily
consistent with our observation.

{\em Polarimetry.}
The synthetic polarization maps of Fischer et~al.\
(\cite{FischerHenningEtAl96}) show that a significant polarization at
the position of the star is only present for nearly edge--on
configurations.  The high degree of polarization of A ($\sim14\%$) is thus
not in agreement with A being the star.  At larger separations from A the
polarization pattern is centrosymmetric.  The center of such a pattern
is thought to be at the position of the illuminating source (Fischer
et~al.\ \cite{FischerHenningEtAl94}, \cite{FischerHenningEtAl96}).  
The fact that in Fig.~\ref{polmap}b this center does not coincide with A
but is located significantly north of it, supports our interpretation that
A is not the star.
\subsection{Is the star at or near the position of B?}
\label{disb}
In the following (a to d) we will argue that the assumption
that the star is at or near
the position of B is consistent with the observations.

(a) The cometary shapes of A in the $H$ and the $J$ images and the
0.79 $\mu$m and 1.06 $\mu$m HST images (Haniff \& Buscher
\cite{HaniffBuscher98}), as well as the polarization data
strongly suggest that A is part of a scattering lobe within a bipolar
structure.  Consistently, A and its southern tails are relatively blue
($H-K$ ranging from 2 to 3.2 in Fig.~\ref{hkcolor}) compared to the
integral color.

(b) The northern components B, C, and D are, on the other hand, rather
red ($H-K\approx4.2$) in comparison with the integral color.  This
suggests that these structures are strongly obscured and reddened by
circumstellar dust.

(c) The brightest northern component in the $J$ image
(Fig.~\ref{recneb}a) can hardly be seen in the $H$ image and is thus
very blue.
It has a separation of $\sim$500~mas from A at a position angle of
$\sim$27\degr.  In the $K$--band image from the same epoch
(Fig.~\ref{recall}b) component B is at a separation of $\sim$200~mas
and at a position angle of 21\degr.  This means that B is almost in
the center between the northern {\sf X}--shaped $J$-band component
and the southern cometary component A (see Sect.~\ref{jbipolar}).
This morphology suggests that
the latter components are the opposite
lobes of a bipolar structure around a  central star
approximately located at the position of B.

(d) The polarization map (Fig.~\ref{polmap}b) fits well with the
picture that the star is at B.
The region between the two $J$--band lobes is only weakly polarized
with polarization null points in the east and northwest.  Such null
points indicate a transition from an optically thick region with
multiple scattering to a region with predominantly single scattering
of photons (Piirola et~al.\ \cite{PiirolaScaltritiEtAl92}, Fischer
et~al.\ \cite{FischerHenningEtAl94}, \cite{FischerHenningEtAl96}). 
The center of the centrosymmetric
polarization pattern is located between the two
lobes.  Because of
the scatter in the direction of the polarization
vectors it is not possible to determine very precisely if the center
coincides with B, but the polarization pattern is consistent
with this assumption.
We note that at the position of B the
polarized intensity (Fig.~\ref{polmap}c) may be slightly contaminated by
the wings of the HST diffraction pattern associated with the dominant
component A.  The polarimetry of A itself, however, is not affected. 
Since the polarimetric features discussed above and in
Sec.~\ref{disa} are only slightly influenced
by this contamination, the conclusions drawn so far do not change.

{\em Changes in the mass loss rate.}
The change of the shape of component A and the fading of B
can be attributed to an increasing mass loss which is
accompanied by a gradual increase of the optical depth of the dust
shell.  This is most obvious for the later observation epochs,
suggesting an enhanced mass loss since 1997.
A strongly variable mass loss has, in fact, been predicted by
theoretical models treating the dust formation mechanism in the
envelopes of long--period variable carbon stars (Fleischer et~al.\
\cite{FleischerGaugerEtAl92}, Winters et~al.\
\cite{WintersFleischerEtAl94}, \cite{WintersFleischerEtAl95}).  Periods
of this mechanism may be significantly longer than the stellar pulsation
period (Winters et~al.\ \cite{WintersFleischerEtAl95}).
An increasing optical depth of the inner dust shell also results in an
increasing apparent separation of the dust shell structures.  The
apparent motion of the components is thus not solely determined by the
velocity of the dust particles but also by the changes of the optical
properties of the circumstellar dust.

{\em Alternative model: the star between A and B.}
From the present observational data it is not possible to exclude
the possibility that
the star may be located between A and B, close to B, or in the center of
A, B, C, and D.  In particular, the precision of the polarization map is
not sufficient to conclude whether the star is at the position of B or
only close to it.

{\em Radiative transfer modeling.}
An answer to the question of where the star is located
requires radiative transfer calculations.
In a second paper (Men'shchikov et~al., in preparation), we will present
the results of our two--dimensional radiative transfer modeling showing
that the shapes of A and B cannot be reproduced when assuming a
position of the star between A and B.  On the other hand, it was
possible to reproduce these shapes and the intensity ratio of A and B in
the case where the star is assumed to be at the position of B.  Clear
preference is thus given to the latter model.

\subsection{Stellar evolution and bipolar structure}
\label{disevol}
IRC\,+10\,216 is without doubt in a very advanced stage of its AGB
evolution due to its long pulsational period, high mass-loss rate, and
carbon-rich dust-shell chemistry indicating that a significant
number of thermal pulses have already taken place.
The star's initial mass can be estimated
to be $4 $M$_{\odot} \pm 1 $M$_{\odot}$ due to the observed isotopic
ratios of C, N and O in the dust shell (Guelin et~al.\
\cite{GuelinForestiniEtAl95}) and the luminosity of the central star
(Weigelt et~al.\ \cite{WeigeltBalegaEtAl98}).  Accordingly, the core
mass should be $\sim 0.7$ to $ 0.8 $M$_{\odot}$ with corresponding
thermal-pulse cycle times of $\sim 1-3\,10^{4}$yr (Bl\"ocker
\cite{Bloecker95}).  Introducing the mean observed mass-loss rate to
these thermal-pulse periods shows that the present stellar wind leads to
a very effective erosion of the envelope per thermal pulse cycle,
possibly as high as $\sim 1 $M$_{\odot}$/cycle.  Consequently, the whole
envelope may be lost during the next few thermal pulses leading to the
termination of the AGB evolution.
Thus, it is not unlikely to assume that IRC\,+10\,216 has entered a
phase immediately before moving off the AGB.  This is strongly supported
by the non-spherical appearence of its dust shell showing even bipolar
structures.  Unlike AGB stars, post-AGB objects as
protoplanetary nebulae often expose prominent features of
asphericities, in particular in axisymmetric geometry (e.g.\ Olofsson
\cite{Oloff93}, \cite{Oloff96}).
Accordingly,
\object{IRC\,+10\,216} can be thought to be an object in transition.  It
is noteworthy that the establishment of bipolar structures, i.e.  the
metamorphosis into a protoplanetary nebulae, obviously already begins
during the (very end of) AGB evolution.
The clumpiness within the bipolar shape is probably due to small scale
fluctuations of the dust condensation radius which, in turn, might be
influenced by, e.g., giant surface convection cells (Schwarzschild
\cite{Schwarzschild75}).  The formation of giant convection cells can be
assumed to be a common phenomenon in red giants.

The shaping of planetary nebulae can successfully be described by
interacting stellar wind models (Kwok et al.\ \cite{Kwoketal78}, Kahn \&
West \cite{KahnWest85}).  Within this scenario a fast (spherical) wind
from the central star interacts with the slow wind of the preceding AGB
evolution.  The slow AGB wind is asssumed to be non-spherical
(axisymmetric) which leads to the observed morphology of planetary
nebulae (Mellema \cite{Mel96}).

The cause of an aspherical AGB mass loss is still a matter of debate.
Different mechanisms to provide the required equatorial density
enhancements are discussed (cf.\ Livio \cite{Liv93}).  Among these,
binarity is one channel including common envelope evolution and spin up
of the AGB star due to the interaction with its companion (Morris
\cite{Mor81}, Bond \& Livio \cite{BondLiv90}).  Not only stellar
companions are found to be able to spin up the AGB star but also
substellar ones as brown dwarfs and planets, most effectively by
evaporation in the AGB star's envelope (Harpaz \& Soker \cite{SokHar92},
Soker \cite{Soker97}).
Currently there is no observational evidence for a possible binary nature of
\object{IRC\,+10\,216}.  The fact that the polarization pattern in the
southeastern part of the nebula at 1.1~$\mu$m has a different
orientation than in the rest of the nebula might be 
an indication of a second illuminating source.

Mechanisms inherent to the star include rotation (Dorfi \& H\"ofner
\cite{DorHoef96}, Garcia-Segura et al.\ \cite{GarBer99}), non-radial
pulsations (Soker \& Harpaz \cite{SokHar92}) and magnetic fields (Pacoli
et al.\ \cite{Pacetal92}, Garcia-Segura et al.\ \cite{GarBer99}).  Both
non-radial $p$-modes and magnetic fields appear to only be important for
significant rotation rates. Often spin-up agents due to
binarity are assumed.  For instance, Groenewegen (\cite{Groenewegen96})
favours non-radial pulsation or an as yet unidentified companion which spun
up the central star as the most likely explanation for the non-spherical
shape of the dust shell of \object{IRC\,+10\,216}.

AGB stars are known to be slow rotators.  Stars with initial masses
below $\sim 1.3 $M$_{\odot}$ can be expected to lose almost their
entire angular momentum during the main sequence phase due to magnetic
braking operating in their convective envelopes.  Consequently, they are
not believed to devolop non-spherical mass-loss due to rotation.
Stars with larger initial masses are spun down due to expansion and mass loss
in the course of evolution, 
but may achieve sufficiently high rotation rates at the end of their AGB
stage (Garcia-Segura et al.\ \cite{GarBer99}). Even small
rotation rates influence dust-driven winds considerably, yielding a mass
loss preferentially driven in the equatorial plane (Dorfi \& H\"ofner
\cite{DorHoef96}).  Furthermore, for supergiants leaving the Hayashi
line, Heger \& Langer (\cite{HegLan98}) found that significant spin up of
the surface layers may take place.  Thus, on second glance, inherent
rotation might
be able to support axisymmetric mass loss during the transition to
the proto-planetary nebula phase for more massive AGB stars
as \object{IRC\,+10\,216}.

\section{Conclusions}
We have presented high-resolution $J$--, $H$--, and $K$--band observations
of \object{IRC +10 216} with the highest resolution so far at $H$ of
70~mas.  A series of $K$--band images from five epochs between October
1995 and November 1998 shows that the inner nebula is non-stationary.
The separations of the four dominant resolved components increased within
the 3 years by up to $\sim35\%$.  For the two brightest components a relative
velocity within the plane of the sky of about 23~mas/yr or 14~km/s was
found.  Within these 3 years the rather faint components C and
D became brighter whereas component B faded.  The general geometry
of the nebula seems to be bipolar.

We find that the most promising model to explain the structures and
changes in the inner nebula is to assume that the star is at or near the
position of component B.  The star is then strongly but not totally obscured
at $H$ and $K$.  Consistently component B is very red in the $H-K$ color
while A and the northern $J$--band components are relatively blue. 
The polarization pattern with strong polarization in the northern arms and
also a significant polarization in the peak supports this model. The
inner nebula and the apparent motions seem to be rather symmetric around
this position and the observed changes are consistent with the
assumption of an enhanced mass loss becoming apparent at least in 1997.

\object{IRC\,+10\,216} is without doubt in a very advanced stage of its
AGB evolution. The observed bipolarity of its dust shell even reveals that
it has possibly entered the phase of transformation into a protoplanetary
nebula.

\begin{acknowledgements}
We thank the referee for valuable comments. 
This research has made use of the SIMBAD database, operated at CDS,
Strasbourg, France.  The HST images for the polarization
analysis have been retrieved from the Hubble Data
Archive operated at the STScI, Baltimore, USA.
\end{acknowledgements}
\end{document}